\documentclass[twocolumn]{aastex631}

\newcommand{\teff}{$T_{\rm eff}$}
\newcommand{\logg}{$\log{g}$}
\newcommand{\feh}{[Fe/H]}
\newcommand{\ALi}{$\textit{A}(\mathrm{Li})$}
\newcommand{\snrr}{$\rm S/N_r$}

\shorttitle{LAMOST Li-rich giants}
\shortauthors{Ding et al.}
\submitjournal{ApJS}
\graphicspath{./}
\accepted{2026.02.17}
\begin{document}

\title{Identify $\sim$20,000 Li-rich Giants in the LAMOST Low-Resolution Survey}
\correspondingauthor{Liang Wang, Jian-Rong Shi, Hong-Liang Yan} 
\email{liangwang@niaot.ac.cn, sjr@nao.cas.cn, hlyan@nao.cas.cn}

\author[0000-0001-6898-7620]{Ming-Yi Ding}
\affiliation{Nanjing Institute of Astronomical Optics \textnormal{\&} Technology, Chinese Academy of Sciences, Nanjing 210042, China}
\affiliation{CAS Key Laboratory of Astronomical Optical \textnormal{\&} Technology, Nanjing Institute of Astronomical Optics \textnormal{\&} Technology, Nanjing 210042, China}

\author[0000-0003-3603-1901]{Liang Wang}
\affiliation{Nanjing Institute of Astronomical Optics \textnormal{\&} Technology, Chinese Academy of Sciences, Nanjing 210042, China}
\affiliation{CAS Key Laboratory of Astronomical Optical \textnormal{\&} Technology, Nanjing Institute of Astronomical Optics \textnormal{\&} Technology, Nanjing 210042, China}
\affiliation{University of Chinese Academy of Sciences, Beijing 100049, China}

\author[0000-0002-0349-7839]{Jian-Rong Shi}
\affiliation{CAS Key Laboratory of Optical Astronomy, National Astronomical Observatories, Chinese Academy of Sciences, Beijing 100101, China}
\affiliation{School of Astronomy and Space Science, University of Chinese Academy of Sciences, Beijing 100049, China}

\author[0000-0002-8609-3599]{Hong-Liang Yan}
\affiliation{CAS Key Laboratory of Optical Astronomy, National Astronomical Observatories, Chinese Academy of Sciences, Beijing 100101, China}
\affiliation{School of Astronomy and Space Science, University of Chinese Academy of Sciences, Beijing 100049, China}

\author[0000-0002-6647-3957]{Chun-Qian Li}
\affiliation{School of Physics and Astronomy, Beijing Normal University, Beijing 100875, China}
\affiliation{Institute for Frontiers in Astronomy and Astrophysics, Beijing Normal University, Beijing 102206, China}

\author[0000-0001-8869-653X]{Jing Chen}
\affiliation{Nanjing Institute of Astronomical Optics \textnormal{\&} Technology, Chinese Academy of Sciences, Nanjing 210042, China}
\affiliation{CAS Key Laboratory of Astronomical Optical \textnormal{\&} Technology, Nanjing Institute of Astronomical Optics \textnormal{\&} Technology, Nanjing 210042, China}

\author[0000-0001-9788-6378]{Kai Zhang}
\affiliation{Nanjing Institute of Astronomical Optics \textnormal{\&} Technology, Chinese Academy of Sciences, Nanjing 210042, China}
\affiliation{CAS Key Laboratory of Astronomical Optical \textnormal{\&} Technology, Nanjing Institute of Astronomical Optics \textnormal{\&} Technology, Nanjing 210042, China}
\affiliation{University of Chinese Academy of Sciences, Beijing 100049, China}

\author{Yu-Lu Liang}
\affiliation{School of Physics and Electronic Science, Guizhou Normal University, Guiyang 550025, China}

\author[0009-0006-0874-3273]{Zi-Chong Zhang}
\affiliation{School of Astronomy and Space Science, Nanjing University, Nanjing 210093, China}
\affiliation{Key Laboratory of Modern Astronomy and Astrophysics, Nanjing University, Ministry of Education, Nanjing 210093, China}

\author{Xu-Hang Yin}
\affiliation{Nanjing Institute of Astronomical Optics \textnormal{\&} Technology, Chinese Academy of Sciences, Nanjing 210042, China}
\affiliation{CAS Key Laboratory of Astronomical Optical \textnormal{\&} Technology, Nanjing Institute of Astronomical Optics \textnormal{\&} Technology, Nanjing 210042, China}

\author[0009-0000-6006-1073]{Qin-Yang Dong}
\affiliation{Nanjing Institute of Astronomical Optics \textnormal{\&} Technology, Chinese Academy of Sciences, Nanjing 210042, China}
\affiliation{CAS Key Laboratory of Astronomical Optical \textnormal{\&} Technology, Nanjing Institute of Astronomical Optics \textnormal{\&} Technology, Nanjing 210042, China}

\begin{abstract}
Li-rich giants serve as valuable tracers of stellar evolution and surface enrichment processes, for which a statistically large and homogeneous sample is crucial. Using the massive low-resolution ($R \sim 1800$) spectroscopic dataset from the Large Sky Area Multi-Object Fiber Spectroscopic Telescope (LAMOST) survey, we systematically search for Li-rich giants by determining their stellar lithium abundances through template matching of the \ion{Li}{1} 6708\,\AA\ absorption line with a grid of synthetic spectra generated from ATLAS9 model atmospheres. The derived abundances are validated against previous high-resolution studies, showing good consistency with a mean absolute error of about 0.15\,dex. Therefore, we adopt a threshold of $A(\rm{Li}) > 1.65$\,dex to select Li-rich candidates, followed by visual inspection to ensure the reliability of each detection. Eventually, more than 20,000 Li-rich giants are identified, corresponding to approximately 2.5\% of all giants in LAMOST DR9. We also investigate the occurrence rate of Li-rich giants in different evolutionary stages. This work presents a large and homogeneous catalog of Li-rich giants derived from the LAMOST low-resolution survey, which provides a reliable and valuable dataset for future studies of stellar evolution and lithium enrichment in evolved stars.
\end{abstract}


\section{Introduction} \label{sec:intro}
Lithium ($\text{Li}$) occupies a unique and significant place in astrophysics, serving as a critical tracer of both Big Bang nucleosynthesis and stellar internal processes. The primordial abundance of ${^7\text{Li}}$ is established shortly after the Big Bang, providing a fundamental cosmological constraint \citep{Spite1982AbundanceLithiumUnevolved, Fields2020BigBangNucleosynthesisPlanck}. Lithium is extremely fragile, undergoing destruction through proton capture reactions ($\text{p} + {^7\text{Li}} \rightarrow 2 \ {^4\text{He}}$) at relatively low temperatures, typically above $2.5 \times 10^6$\,K. Therefore, the surface lithium abundance in stars acts as a sensitive probe of internal mixing and stellar evolutionary stage. During the MS phase, stars with relatively shallow convective envelopes preserve a substantial fraction of their initial surface lithium. However, as a star evolves towards the Red Giant Branch (RGB), the convective zone becomes deeper and transports surface material to the hotter interior during the first dredge-up, where ${^7\text{Li}}$ is efficiently destroyed. This process predicts a sharp and dramatic decrease in surface lithium abundance, \ALi, for RGB stars, with a threshold of $A\rm(Li) < 1.5$\,dex \citep{Iben1967StellarEvolutionVI, Charbonnel2000NatureLithiumRich, Kirby2016LITHIUMRICHGIANTSGLOBULAR, Charbonnel2020LithiumRedGiant}. Although the vast majority of giants agree to this prediction, a small fraction exhibits anomalously high $A(\text{Li})$ values, far exceeding the theory prediction, which are categorized as Li-rich giants. The first such object identified is a K-giant with an extreme lithium abundance ($A\rm(Li)\sim 3.2$\,dex) reported by \citet{Wallerstein1982GiantUnusuallyHigh}, and numerous other Li-rich giants have been reported across various Galactic components. Observations demonstrated that lithium enrichment occurs across a vast range of metallicities, including extremely metal-poor stars with \feh $\sim -4.0$\,dex \citep{Mucciarelli2024TrueNatureHE00575959}. Furthermore, Li-rich giants can be found at any evolutionary stage \citep{Kirby2016LITHIUMRICHGIANTSGLOBULAR, Li2018EnormousLiEnhancement} and have also been reported in dwarf galaxies \citep{Kirby2012DiscoverySuperLiRich}. The occurrence of these objects presents a long-standing puzzle in stellar evolution, indicating that nonstandard processes are needed to explain the observed surface lithium enrichment \citep{Lagarde2010LiSurveyGiant, Casey2016GaiaESOSurveyRevisiting}.

Theoretical explanations for the Li rich phenomenon generally fall into two primary categories: internal production and external pollution. The internal production mechanisms invoke the Cameron-Fowler (CF) mechanism \citep{Cameron1971LithiumSPROCESSRedGiant}, in which ${^7\text{Be}}$ is synthesized in the hot interior layers by ${^3\text{He}(\alpha,\gamma)^7\text{Be}}$ reaction and subsequently transported to the cooler envelope regions where the electron capture converts it to observable ${^7\text{Li}}$ before it can be destroyed. However, implementing the CF mechanism in evolved, low-mass stars requires nonstandard extramixing processes that operate on timescales short enough to preserve the ${^7\text{Be}}$ during transport. The triggers for this extramixing include rotational driven mixing \citep[e.g.,][]{Charbonnel2010ThermohalineInstabilityRotationinduced, Charbonnel2000NatureLithiumRich, Denissenkov2010MODELMAGNETICBRAKING, Tsantaki2023SearchLithiumrichGiants}, thermohaline instability \citep{Charbonnel2000NatureLithiumRich, Charbonnel2007ThermohalineMixingPhysical}, magnetically induced mixing \citep{Busso2007CanExtraMixing, Nordhaus2008MagneticMixingRed, DeLaReza2025LithiumrichGiantStars} and internal gravitational wave induced mixing \citep{Wu2025PhysicalMechanismFormation}; each can connect the deep production layer to the convective envelope. Furthermore, recent asteroseismic investigations have provided refined insights, suggesting that the core helium flash event at the RGB tip is likely a key evolutionary phase that facilitates this internal Li enrichment \citep{Kumar2020DiscoveryUbiquitousLithium, Zhang2021LithiumEvolutionGiant, Singh2021TrackingEvolutionLithium}. External pollution mechanisms focus on mass transfer from a Li-rich external source to the envelope \citep[e.g.,][]{Li2025EffectMatterAccretion}. The proposed routes include accretion from a binary companion \citep{Singh2025LienrichmentRedClump, Sayeed2025LookingCompanionshipRadial, Sayeed2025ProbingBinaryArchitectures}, tidal disruption \citep{Denissenkov2004EnhancedExtraMixing, Casey2019TidalInteractionsBinary}, merger of a stellar or substellar companion \citep{AguileraGomez2016LithiumrichRedGiants, Zhang2020PopulationSynthesisHelium}, and planet engulfment \citep[e.g.,][]{Siess1999AccretionBrownDwarfsa, Carlberg2009RolePlanetAccretion}. Although these external scenarios can provide Li-enrichment to some extent, they are often insufficient to fully explain certain observed phenomena, specifically the population of Li-rich stars that gather in the Red Clump (RC) phase \citep{Yan2020MostLithiumrichLowmass} and the existence of the most extreme super Li-rich giants. Furthermore, there are also theories that explain this phenomenon by considering suppression of main-sequence lithium depletion \citep[e.g.,][]{Li2024ConvectiveMixingFormation}. Consequently, the lithium enrichment of giants remains a complex and challenging issue in stellar astrophysics, likely resulting from a confluence of multiple internal and external nonstandard mechanisms acting in combination.

Large spectroscopic surveys have recently revolutionized the study of Li-rich giants. Programs such as the {\it Gaia}-ESO Survey \citep{Gilmore2012GaiaESOPublicSpectroscopic}, the GALactic Archeology with HERMES survey \citep[GALAH,][]{DeSilva2015GALAHSurveyScientific}, the Large Sky Area Multi-Object Fiber Spectroscopic Telescope \citep[LAMOST,][]{Cui2012LargeSkyArea} and the Sloan Digital Sky Survey \citep[SDSS,][]{Eisenstein2011SDSSIIIMassiveSpectroscopic} have provided massive spectral data for statistical studies. The {\it Gaia}-ESO Survey, using UVES and GIRAFFE spectrographs on the Very Large Telescope (VLT), provided high-resolution optical spectra that identified dozens of Li-rich giants \citep{Casey2016GaiaESOSurveyRevisiting, Smiljanic2018GaiaESOSurveyProperties, Franciosini2022GaiaESOSurveyLithiuma, Nepal2023GaiaESOSurveyPreparing}. 
Utilizing the seventh SDSS data release (DR7), \citet{Martell2013LithiumrichFieldGiants} identified 27 Li-rich field giants based on follow-up high-resolution spectra.
The GALAH survey, which employs the HERMES spectrograph to derive high-resolution ($R \sim 28,000$) spectra \citep{Martell2017GALAHSurveyObservational}, has made significant progress. \citet{Deepak2019StudyLithiumRich} reported the discovery of 335 new Li-rich giants from GALAH DR2 \citep{Buder2018GALAHSurveySecond}; however, their subsequent investigations did not find a clear link between high Li abundance and rotational or tidal interaction in that subset \citep{Deepak2020AbundanceAnalysesLienriched}. Based on the GALAH DR3 dataset \citep{Buder2021GALAHSurveyThird}, the mechanisms of lithium enrichment in the giant phase have been extensively investigated by recent works \citep{Martell2017GALAHSurveyObservational, Chaname2022MassMattersNo, CastroTapia2024AreLithiumRichGiants, Sayeed2024ManyRoadsLead}. The LAMOST low-resolution survey (LRS) has dramatically expanded the sample size to tens of thousands \citep{Casey2019TidalInteractionsBinary, Gao2019LithiumrichGiantsLAMOST, Cai2022LirichGiantsIdentified}, providing a vast foundation upon which to demonstrate the diverse physical properties of Li-rich giants through further dedicated investigations \citep{Yan2018NatureLithiumEnrichment, Zhou2018SuperLithiumrichGiant, Zhou2019HighresolutionSpectroscopicAnalysis, Zhou2021LAMOSTHRSSpectroscopic, Zhou2022LirichGiantsLAMOST, DeLaReza2025LithiumrichGiantStars}.
The combination of LAMOST spectroscopy with asteroseismic data from the {\it Kepler} and {\it TESS} missions is particularly helpful, which reveals more detailed evolutionary stages, showing that low-mass Li-rich giants are predominantly located in the RC rather than in the canonical RGB \citep{Yan2020MostLithiumrichLowmass, Kumar2020EvolutionLithiumLowmass, Singh2021TrackingEvolutionLithium}. A series of recent studies have further explored He–flash–induced Li enrichment, specifically spanning the RGB Tip to the core-helium-burning (HeB) phase \citep{Singh2019SurveyLirichGiants, Schwab2020HeliumflashinducedMixingEvent, Kumar2020DiscoveryUbiquitousLithium, Kumar2020EvolutionLithiumLowmass, Zhang2021LithiumEvolutionGiant, Deepak2021LithiumRedGiants, Mallick2023LithiumAbundancesGiants, Mallick2025HighLithiumAbundance}. The combined observational and theoretical efforts fundamentally expanded our understanding of Li-rich giants and imposed stronger constraints on the enrichment mechanisms.

In this work, we expand the census of Li-rich giants using the large and homogeneous stellar sample provided by LAMOST LRS ninth data release (DR9), enabling a new constraint of lithium enrichment across a wide region of parameter space. Section~\ref{sec:obs} describes the LAMOST observations and the adopted stellar parameters. Section~\ref{sec:met} outlines our methodology based on the template matching of \ion{Li}{1} 6708\,\AA\ feature. Section~\ref{sec:res} presents the abundance validation and the manual inspection used to ensure the reliability of the Li-rich candidates selected. In Section~\ref{sec:dis}, we discuss the properties of the resulting catalog and the characteristics of the identified Li-rich giants. A brief summary of our findings is provided in Section~\ref{sec:con}.

\section{Observation and Stellar Parameters} \label{sec:obs}
 LAMOST, located at Xinglong Observatory in Hebei Province, China, is a 4\,m–class reflective Schmidt telescope designed for large-scale spectroscopic surveys. Due to its unique optical design, LAMOST offers a wide field of view ($\sim$5$^{\circ}$) and a highly multiplexed focal plane equipped with 4000 optical fibers that feed 16 spectrographs and 32 CCD cameras (4K $\times$ 4K). This configuration enables the simultaneous acquisition of 4000 spectra in a single exposure, making LAMOST particularly efficient for large-scale Galactic and stellar population studies.  

The LAMOST survey has progressed through multiple annual surveys. The low-resolution mode covers 3690--9100\,\AA\ with a resolution power of $R \sim 1800$ near 5500\,\AA, combining data from a blue (3690--5900\,\AA) and a red (5700--9100\,\AA) CCD channel. Although the medium-resolution mode provides detailed spectral information, the low-resolution mode delivers a significantly larger dataset, which is particularly advantageous for constructing a statistically homogeneous and comprehensive sample of Li-rich giants. The DR9 provides 10.8 million low-resolution spectra for 10.4 million stars.

The stellar atmospheric parameters, including effective temperature ($T_{\mathrm{eff}}$), surface gravity ($\log g$), metallicity ([Fe/H]), and radial velocity (RV), are adopted from the LAMOST Stellar Parameter Pipeline (LASP). LASP derives stellar parameters via a two-stage scheme, utilizing spectra and initial RVs provided by the 1D pipeline. In each stage, Correlation Function Interpolation (CFI) is first employed to obtain initial estimates; these serve as starting guesses for ULySS software\citep{Koleva2009ULySSFullSpectrum}, which then determines the final parameters through $\chi^2$ minimization against the ELODIE spectral library\citep{Prugniel2001DatabaseHighMediumresolution}.

\begin{figure}[!htbp]
	\centering
	\fig{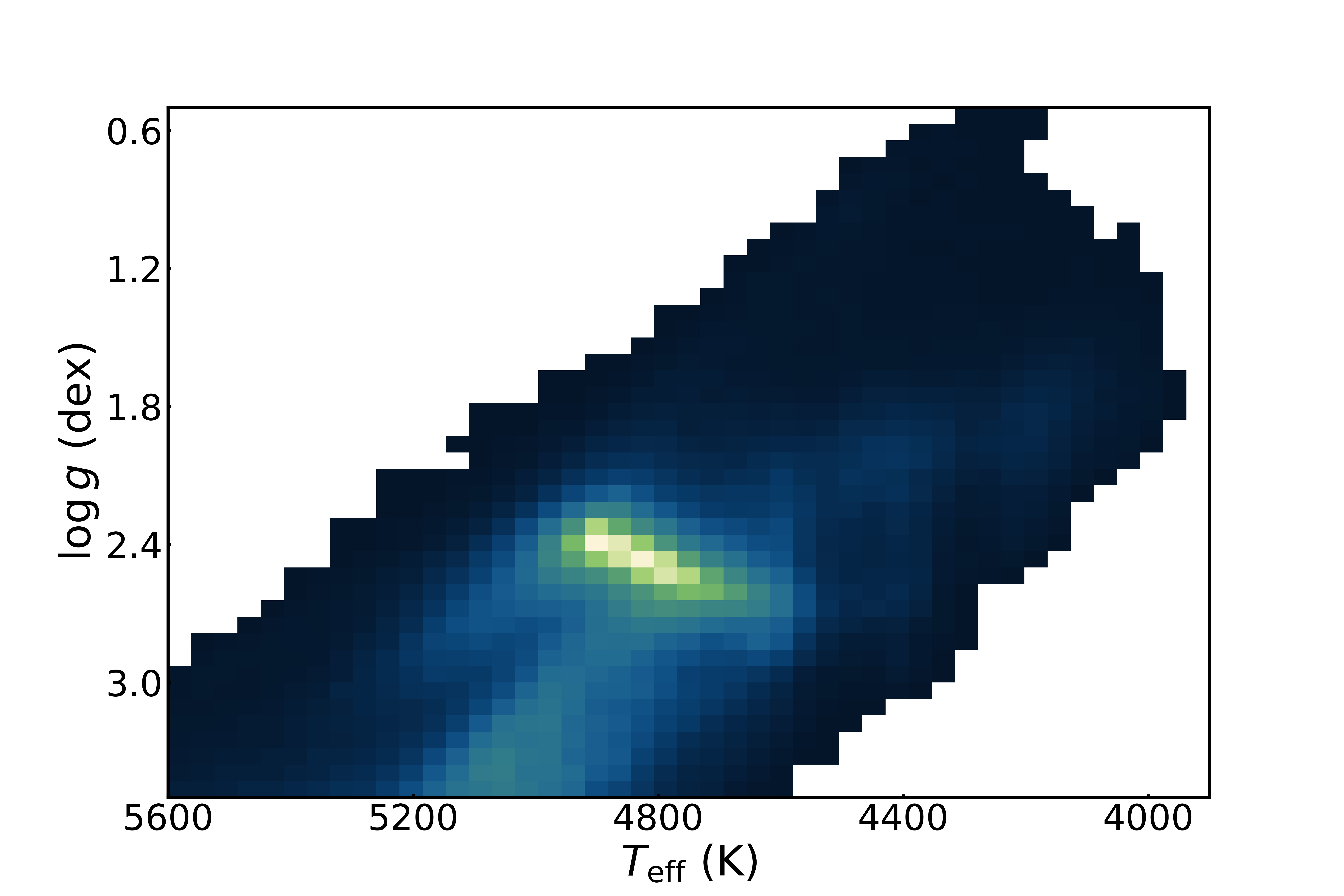}{1\columnwidth}{}
	\caption{HR diagram of the initial giant sample, using stellar parameters from LASP, color-coded by number density.}
	\label{fig:HR_diagram_data}
\end{figure}

For this study, we select 798,594 individual giant stars from LAMSOT LRS DR9, constrain to $3800\,\mathrm{K} < T_{\mathrm{eff}} < 5600\,\mathrm{K}$ and $0.5\,\text{dex} < \log g < 3.5\,\text{dex}$. Figure~\ref{fig:HR_diagram_data} shows the Hertzsprung-Russell (HR) diagram of these stars, illustrating the parameter space of our adopted sample.

\section{Methodology} \label{sec:met}
Lithium abundances in stellar atmospheres can be determined using various spectroscopic techniques, such as equivalent width measurements, curve of growth analysis, spectral synthesis, template matching, and machine learning methods. Here, we employ a template matching approach in which each observed spectrum is compared with a grid of synthetic spectra to derive its lithium abundance and identify Li-rich giant candidates.

\subsection{The Templates}
Based on a template matching method, \citet{Gao2019LithiumrichGiantsLAMOST} identified 10,535 Li-rich giants from LAMOST LRS DR7. We set a similar method as that used in \citet{Gao2019LithiumrichGiantsLAMOST}, which first generates a grid of templates and then interpolates them to compare them with the observed spectrum. For this purpose, we utilize the SPECTRUM synthesis code\footnote{\url{https://www.appstate.edu/~grayro/spectrum/spectrum.html}} to generate the synthetic spectra with the stellar atmosphere models based on ATLAS9 \citep{Castelli2003NewGridsATLAS9}.
The lithium atomic line data are adopted from \citet{Shi2007LithiumAbundancesMetalpoor}.

We first construct a set of synthetic templates covering the wavelength range of 6600 to 6800\,\AA, which includes the \ion{Li}{1} \,6708\,\AA\ feature along with several nearby spectral lines. The templates are degraded to match the actual resolution of each observational spectrum (typically $R \sim 1800$) by convolving with a broadening core, treating the width as a free parameter. The templates are resampled at a step of 1\,\AA\ to match the observations. Although the micro-turbulence velocity ($v_{\text{micro}}$) influences the line profile, we utilize a fixed value of $v_{\text{micro}} = 1.5\,\mathrm{km\,s^{-1}}$ in our analysis, which is a typical value that ensures consistent results without introducing significant errors. 

\begin{deluxetable}{lrr}
	\centering
	\tablecaption{The stellar parameter space of our template grid}
	\label{Tab:grid}
	\tablehead{
	\colhead{Variable} & \colhead{Range} & \colhead{Step}}
	\startdata
		\teff\ (K)      & 3800 -- 5600        & 100      \\
		\logg\ (dex)    & 0.00 -- 4.00        & 0.25       \\
		\feh\ (dex)     & $-2.60$ -- 0.40     & 0.20   \\
        $\rm [Li/Fe]$ (dex) & $-3.00$\,--\,6.10 & 0.10   \\
	\enddata
\end{deluxetable}

The initial grid of templates is computed across a parameter space representative of giant stars, as summarized in Table~\ref{Tab:grid}. For each target, templates are obtained by linearly interpolating within the pre-calculated grid using the stellar parameters ($T_{\mathrm{eff}}$, $\log g$, [Fe/H]) provided by LASP. To determine the lithium abundance, a sequence of interpolated templates with different [Li/Fe] values is then generated for $\chi^2$ minimization. This allows a direct comparison between the observed spectra and model templates, providing an estimation of the best-fitting $A(\mathrm{Li})$ value for each star.

\begin{figure}[!htbp]
	\centering
	\fig{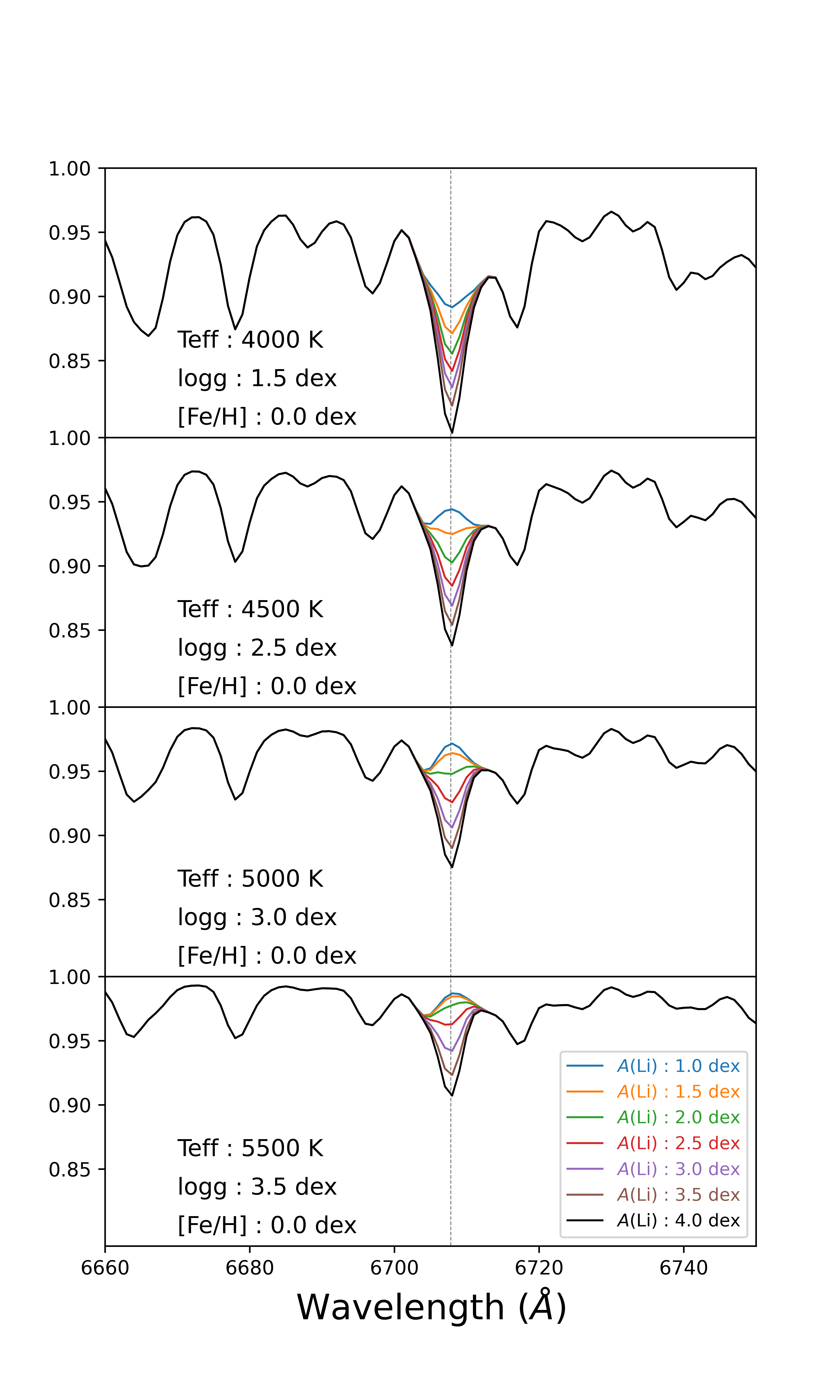}{1\columnwidth}{}
	\caption{Variations of the \ion{Li}{1} 6708\,\AA\ line in synthetic templates with \ALi\ from 1.0 to 4.0\,dex at different effective temperatures.}
	\label{fig:model_var}
\end{figure}

Based on the setup described above, we generate templates that are directly comparable to the observed spectra, enabling the estimation of the best-fitting $A(\mathrm{Li})$ for each star through $\chi^2$ minimization. Figure~\ref{fig:model_var} illustrates the variations of the \ion{Li}{1} 6708\,\AA\ line and the neighboring spectral features with different $T_{\mathrm{eff}}$ values. It highlights how the line strength and profile change with temperature and lithium abundance, and also provides a visual indication of the detection limits for lithium content in giants.

\subsection{The Observed Spectra}
For the LAMOST low-resolution spectra, the wavelength coverage extends from 3690––9100\,\AA, combining data from both the blue and red channels. Each spectrum is first corrected for RV and shifted to the rest frame, and the wavelength is converted from vacuum to air to ensure consistency with the synthetic templates. To prepare the spectra for comparison, we perform flux normalization in the region surrounding the \ion{Li}{1} 6708\,\AA\ absorption line, spanning 6675––6740\,\AA, while masking the core region of 6705––6710\,\AA.

Following the continuum fitting approach of \citet{Li2024LAMALAMOSTMediumResolution}, an iterative local regression fitting is applied using the nearest template to determine the pseudocontinuum. To mitigate edge effects and prevent overfitting, we use a third-order polynomial to fit the pseudocontinuum and remove data points with flux values exceeding $5\sigma$ (typically caused by emission lines or cosmic rays) and those below $-1\sigma$ (associated with random noise or bad pixels). The normalized spectrum is then obtained as $f_{\rm norm} = f_{\rm obs} / f_{\rm continuum}$. This process is repeated iteratively three times to exclude outliers and produce the final normalized spectrum, which is subsequently used for abundance fitting.

\subsection{The \texorpdfstring{$\chi^2$}{chi-squared} Calculation} \label{subsec:chi2}
In order to deduce the level of lithium content from the spectra, our search for Li-rich giant stars mainly focuses on the measurement of the \ion{Li}{1} absorption line at 6708\,\AA, which is located in the red band of the LAMOST low-resolution spectrum. For a celestial target concerned, our $\chi^2$ minimization algorithm is performed on the templates and the observed spectrum. The $\chi^2$ is calculated using the following equation:

$$
\chi^2 = \sum_{i=1}^{m} \frac{(F_i - E_i)^2}{\sigma^2_i}
$$

Here, $F_i$ represents the flux of the template spectrum at the $i_{\text{th}}$ wavelength point, while $E_i$ is the flux of the observed spectrum. The $\sigma_i$ is the measurement error. The $\chi^2$ can be an indicator of how similar a template to the one observed in the wavelength region we consider.

Since the templates are generated from different [Li/Fe], we first calculate a series of $\chi^2$ between the observed spectrum and the templates and searched the minimum $\chi^2$ point using a third-order polynomial regression. Afterwards, we determine the final [Li/Fe] by linear interpolation of the nearest templates, and retrieve the lithium abundance from the formula : $A(\mathrm{Li}) = \mathrm{[Li/Fe]} + \mathrm{[Fe/H]} + A(\mathrm{Li})_{\odot}$.

\begin{figure*}[!htbp]
	\centering
    \gridline{
		\fig{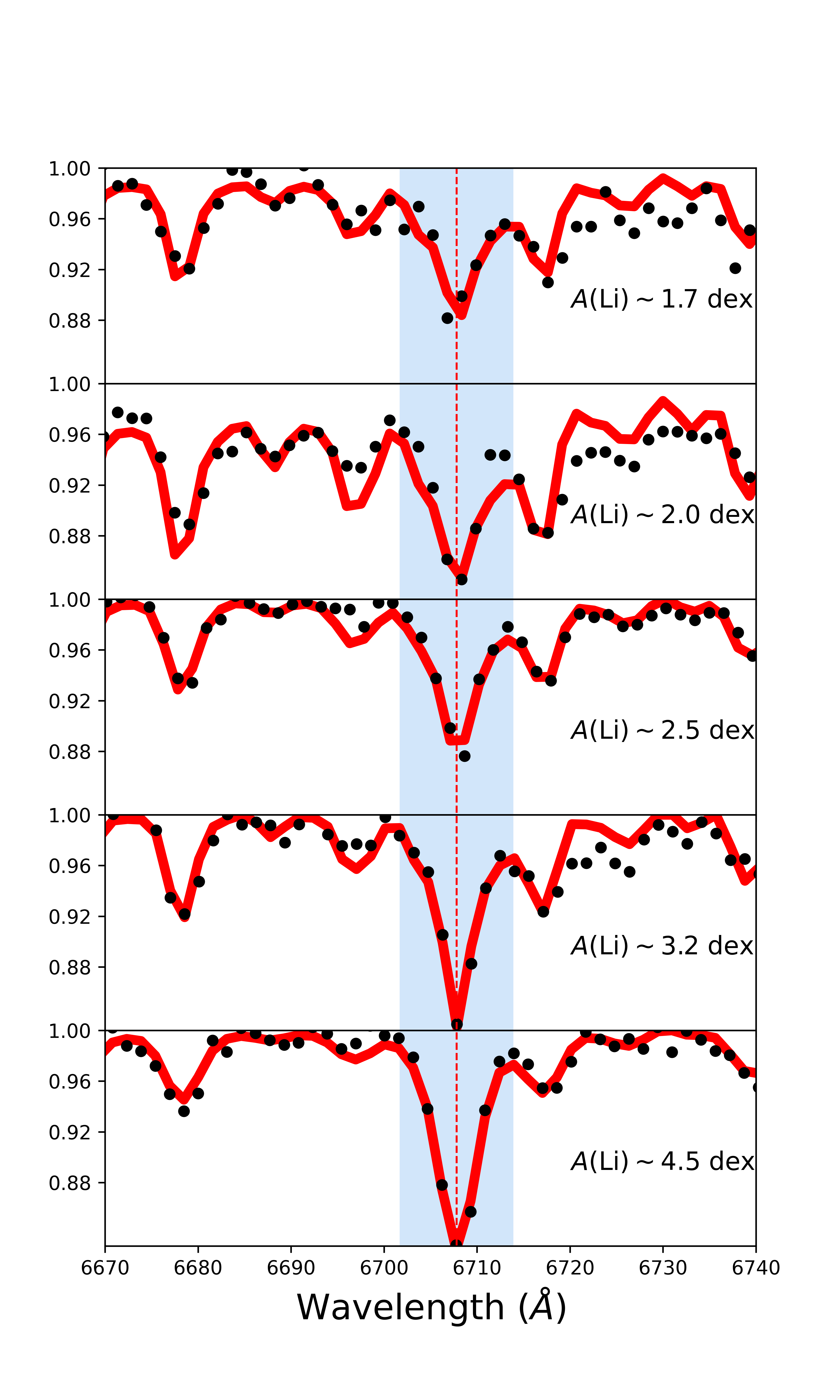}{0.25\textwidth}{}
		\fig{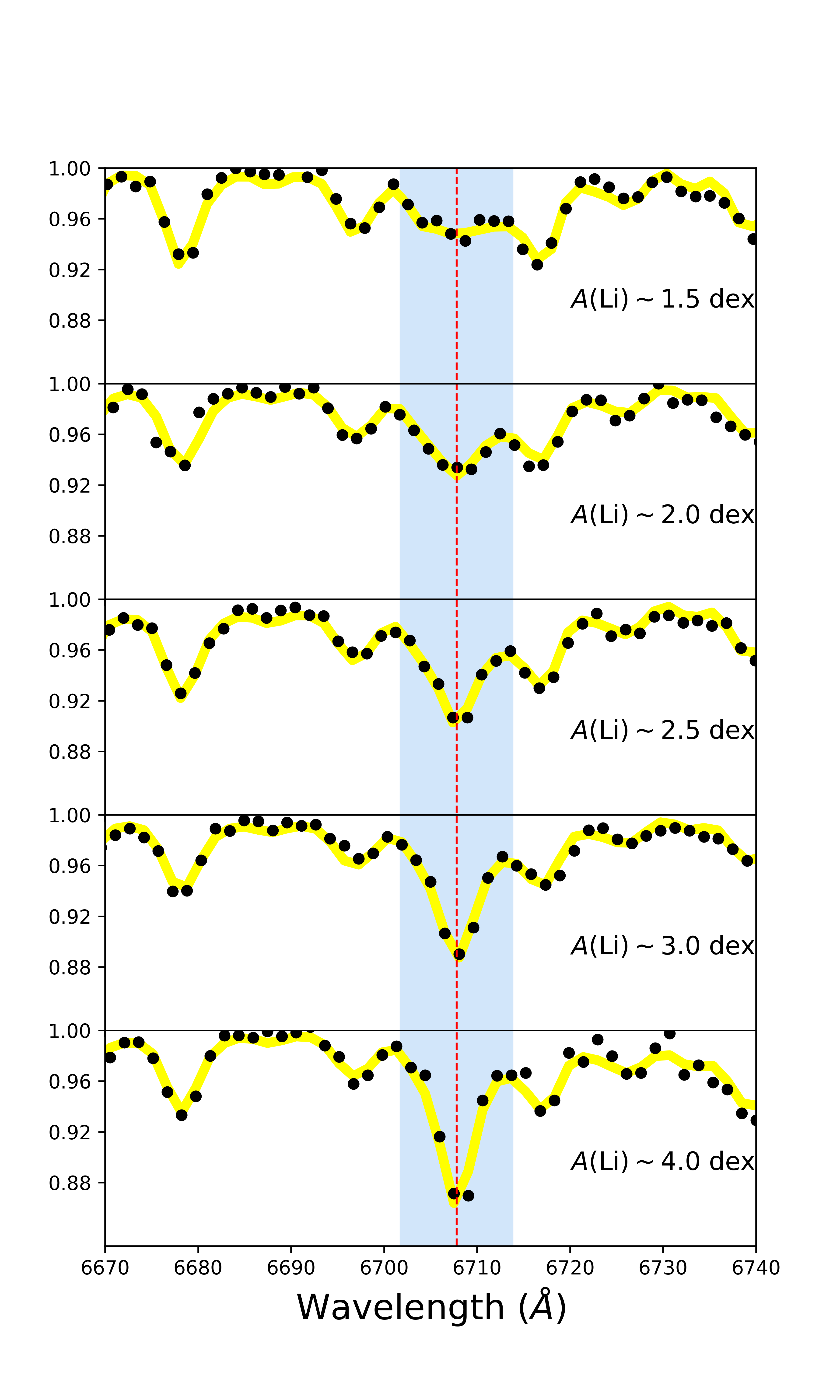}{0.25\textwidth}{}
		\fig{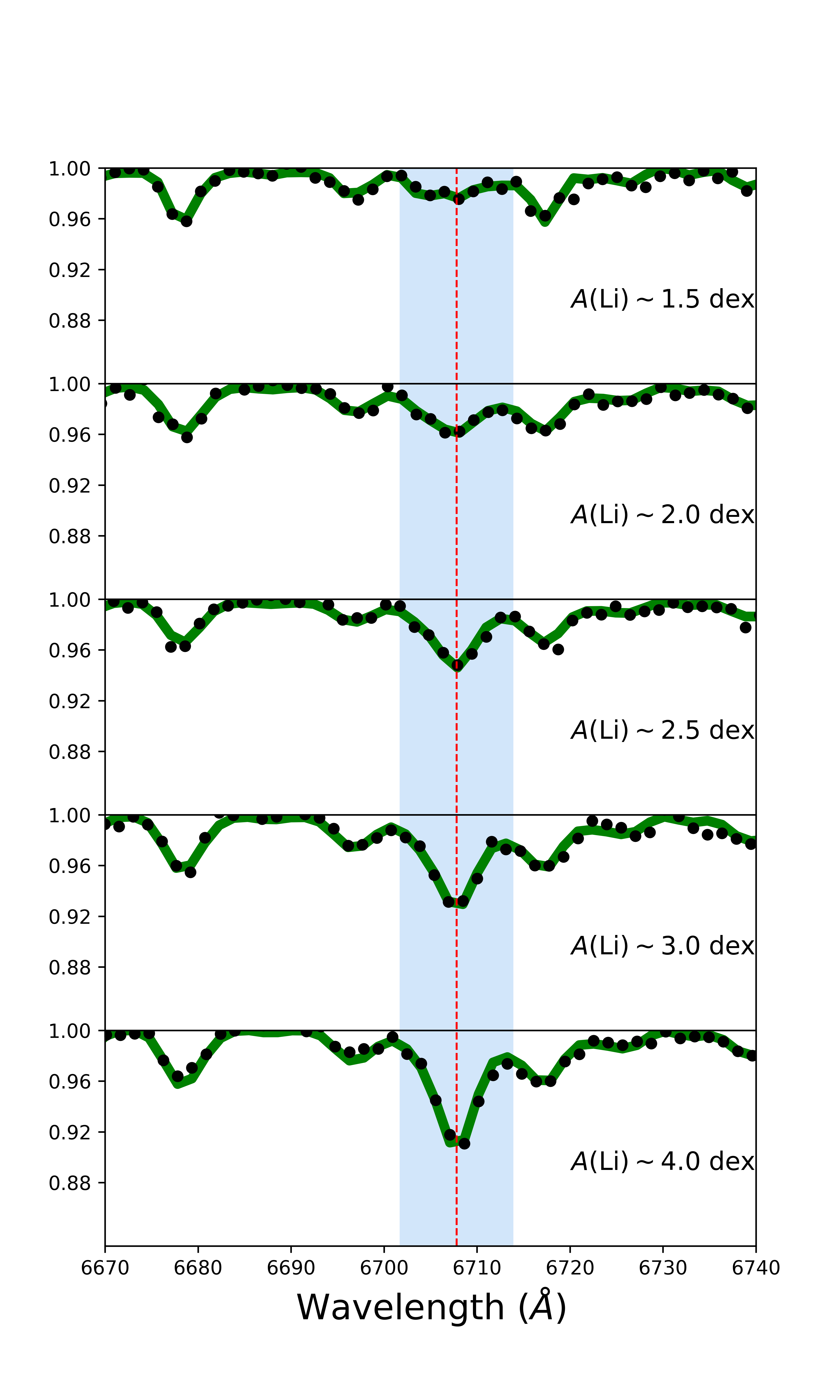}{0.25\textwidth}{}
        \fig{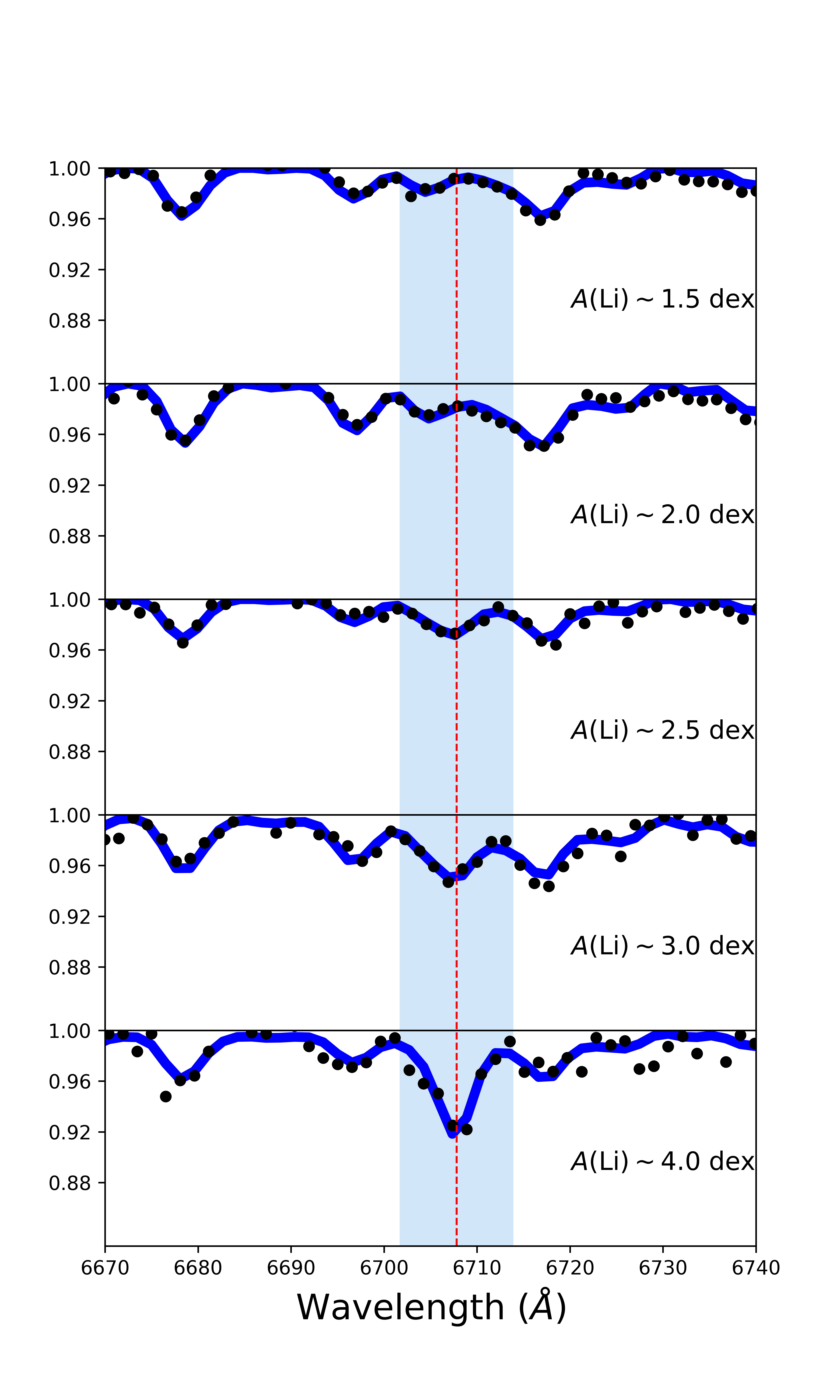}{0.25\textwidth}{}}
	\caption{Variations of the \ion{Li}{1} 6708\,\AA\ line in comparable LAMOST spectra with \ALi\ $\approx$ 1.5––4.0\,dex for different effective temperatures. From left to right: \teff\ $\approx$ 4000, 4500, 5000, and 5500\,K.}
	\label{fig:obs_var}
\end{figure*}

\subsection{The Detection Limits}\label{subsec:limit}
To assess the reliability of our lithium abundance measurements, we need to define the detection limits of our method. Considering that our measurements are based on template matching using low-resolution spectra from the LAMOST survey, the sensitivity to weak \ion{Li}{1} 6708\,\AA\ features naturally varies with stellar parameters, particularly \teff\ and the signal quality in the \ion{Li}{1} 6708\,\AA\ region.

We illustrate this effect in Figure~\ref{fig:obs_var}, where a series of synthetic and observed spectra of different effective temperatures is compared. As shown, the lithium line becomes increasingly weak and less distinguishable at higher temperatures, while in cooler stars the feature remains relatively strong and clearly detectable. This demonstrates that our detection efficiency is highly temperature-dependent: reliable lithium measurements are more easily obtained for cooler giants, whereas for warmer stars the same method may have lower detection efficiency and may fail to distinguish real absorption from spectral noise or blending effects.

To further quantify this limitation, we evaluate the residuals between the observed and best-fitting spectra, along with the local noise level and the measured line depth. When the inferred lithium signal is weaker than either the fitting residuals or the background noise, the result cannot be considered to be an acceptable detection. In these cases, only upper limits of the lithium abundance can be obtained.

By adopting these detection criteria, we ensure that only robust lithium measurements are included in our final sample, while sources with upper limit estimates are excluded from subsequent analysis because their lithium abundance cannot be confidently determined due to data quality or spectral blending.

\section{Result} \label{sec:res}
Using the method described in the previous section, we derive the lithium abundances for 798,594 giant stars with LASP parameters from DR9. Among these, \ALi\ values are successfully estimated for 615,426 stars. In this section, we present the main results of our analysis. We first compare our derived lithium abundances with those reported in previous studies to validate the accuracy of our measurements. Based on these comparisons, we define the criteria to identify and select Li-rich giants from the full sample. Finally, we perform a manual inspection of the selected Li-rich candidates to remove objects that are incorrectly identified by the automated procedure and to ensure that the resulting Li-rich giant sample is highly reliable.

\subsection{Validation with Previous Work}\label{subsec:compare}

\begin{figure}[!htbp]
	\centering
	\fig{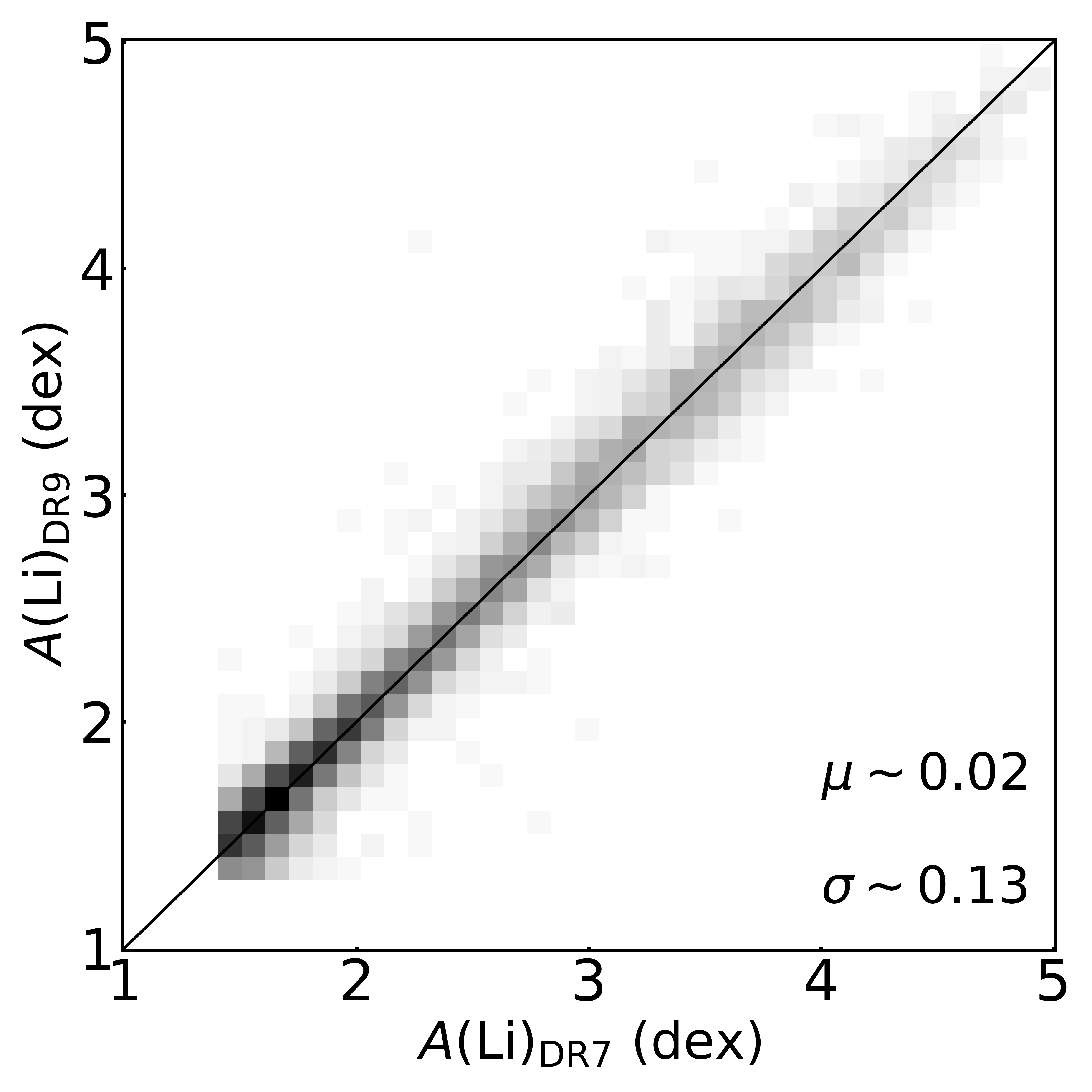}{1\columnwidth}{}
	\caption{Comparison between our derived \ALi\ values and measurements and those from \citet{Gao2019LithiumrichGiantsLAMOST}. The mean difference and standard deviation are $\mu = 0.02$\,dex and $\sigma = 0.13$\,dex, respectively.}
	\label{fig:vsDR7}
\end{figure}

To ensure the reliability of our lithium abundance measurements, it is essential to validate them against external references, particularly those derived from high-resolution spectra. \citet{Gao2019LithiumrichGiantsLAMOST} identified $\sim 10,000$ Li-rich giants from LAMOST DR7 using a template matching approach, demonstrating the potential of low-resolution spectra for conducting a large statistical census of Li-rich stars. In this work, we adopt a similar methodology but implemented with an improved continuum normalization and an updated synthetic template grid. We therefore begin by comparing our derived \ALi\ with those reported by \citet{Gao2019LithiumrichGiantsLAMOST} for the common stars. As shown in Figure~\ref{fig:vsDR7}, the two measurements exhibit excellent agreement, with differences typically below 0.1\,dex, consistent with the uncertainties reported for low-resolution \ALi\ determinations. The remaining small discrepancies likely arise from refinements in our approach.

\begin{figure}[!htbp]
	\centering
	\fig{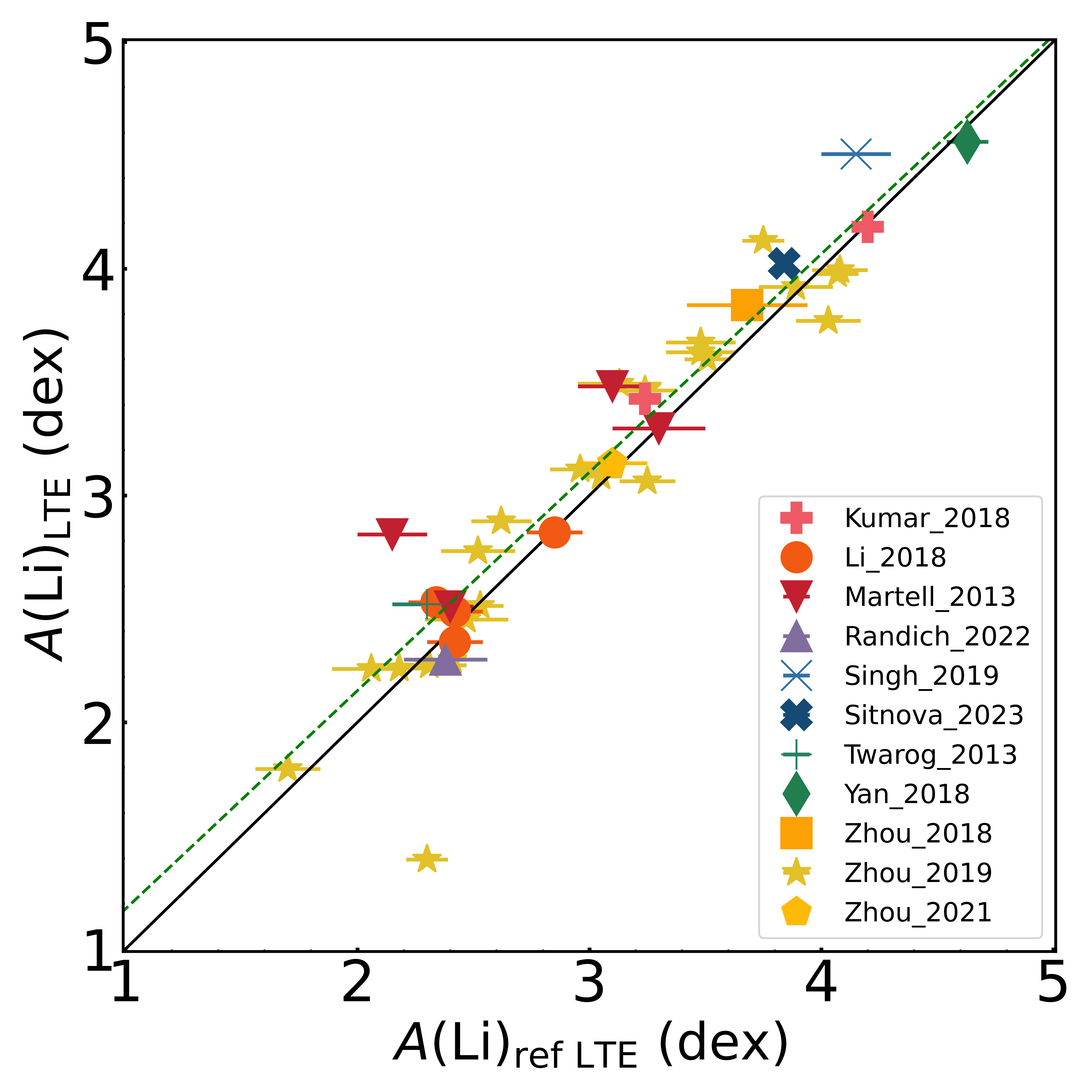}{1\columnwidth}{}
	\caption{Comparisons between our derived \ALi\ values and $A\rm(Li)_{LTE}$ values from high-resolution studies. The green dashed line shows the linear regression.}
	\label{fig:vsHRSlte}
\end{figure}

\begin{figure}[!htbp]
	\centering
    \fig{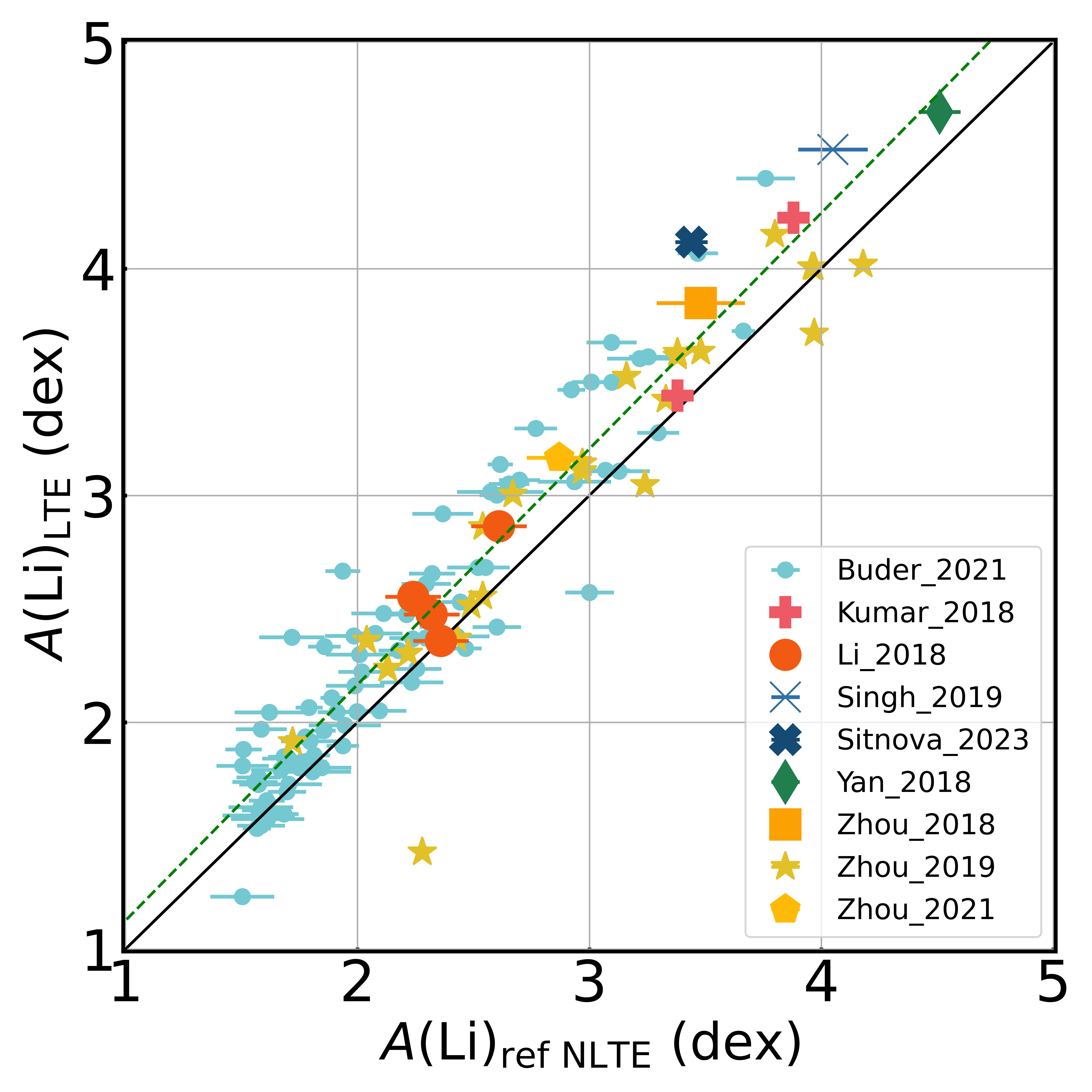}{1\columnwidth}{}
	\caption{Comparisons between our derived \ALi\ values and $A\rm(Li)_{NLTE}$ values from high-resolution studies. The green dashed line shows the linear regression.}
	\label{fig:vsHRSnlte}
\end{figure}

To further validate our measurement, we compare the lithium abundances derived from LAMOST low-resolution spectra with those reported in previous high-resolution studies. We collect 132 common stars from 12 sources in the literature \citep{AnthonyTwarog2013LithiumrichRedGiant, Martell2013LithiumrichFieldGiants, Kumar2018TwoNewSuper, Li2018EnormousLiEnhancement, Yan2018NatureLithiumEnrichment, Zhou2018SuperLithiumrichGiant, Zhou2019HighresolutionSpectroscopicAnalysis, Singh2019SpectroscopicStudyTwo, Buder2021GALAHSurveyThird, Zhou2021LAMOSTHRSSpectroscopic, Randich2022GaiaESOPublicSpectroscopic, Sitnova2023PristineSurveyXXII}, for which local thermodynamic equilibrium (LTE) and/or non-local thermodynamic equilibrium (NLTE) \ALi\ values were derived from high-resolution spectra. The comparison results are presented in Figure~\ref{fig:vsHRSlte} and Figure~\ref{fig:vsHRSnlte}.

For each common star, the reference values ($A\rm(Li)_{ref \ LTE}$, $A\rm(Li)_{ref \ NLTE}$) are taken directly from the literature, while the corresponding $A\rm (Li)_{LTE}$ are recalculated using LAMOST spectra based on the same stellar parameters provided by those studies. Each case is manually inspected, and several mismatched or invalid fits have been excluded to ensure a reliable and accurate comparison.

Figure \ref{fig:vsHRSlte} presents the comparison of the LTE results for 40 stars in common, showing good consistency as expected, since our measurements also assume LTE. The results agree well across the entire \ALi\ range, with a mean absolute error of 0.15\,dex. Figure~\ref{fig:vsHRSnlte} shows the comparison between our LTE-derived \ALi\ values and the NLTE abundances reported for 116 common stars in the literature. Overall, our results are in good agreement with the high-resolution measurements, showing a small systematic overestimation of about 0.22\,dex. A slight increase in dispersion at higher \ALi\ levels can be attributed to the larger NLTE corrections toward high lithium abundance objects.

In the comparison, an outlier is identified, \textit{2MASS J15433057+0030389}. This object is a cool M giant \citep{Li2023ValueaddedCatalogMgiant} whose spectral fit appears acceptable upon visual inspection, yet its derived \ALi\ is underestimated by about 0.8\,dex. Given that this discrepancy likely arises from the complex spectral features of such cool, evolved stars rather than from a fitting error, we retain this source in our comparison sample.

\begin{figure}[!htbp]
	\centering
	\fig{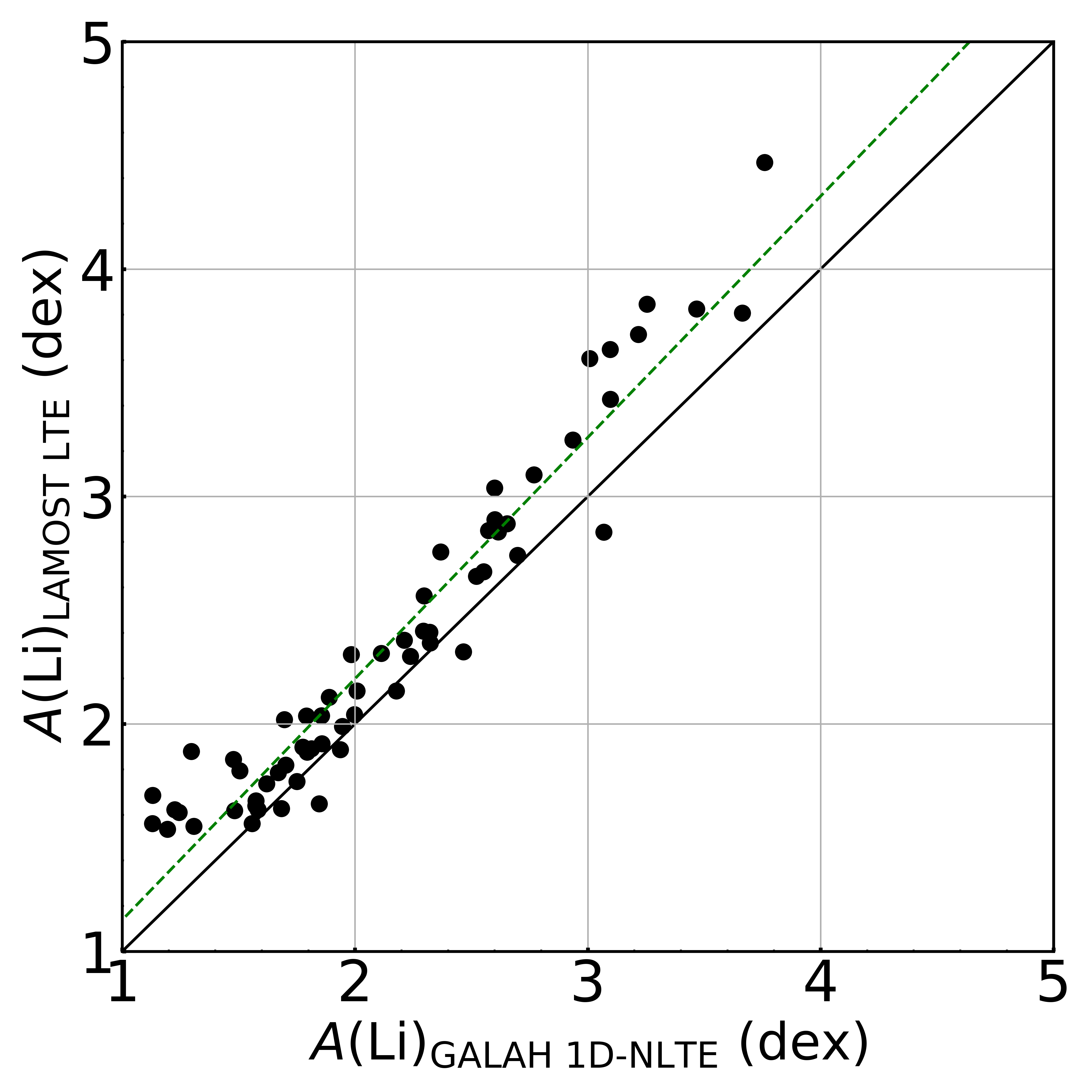}{1\columnwidth}{}
	\caption{Comparison between our derived \ALi\ values and measurements and those 1D-NLTE \ALi\ from the GALAH DR3}. The green dashed line shows the linear regression.
	\label{fig:vsGALAH}
\end{figure}

The GALAH survey has published stellar parameters and chemical abundances for a vast number of stars. Although 3D-NLTE $A(\rm{Li})$ estimates based on machine learning method \citep{Wang20243DNLTELithiuma} are available in DR4 \citep{Buder2025GALAHSurveyData}, we would like to adopt the 1D-NLTE measurements from GALAH DR3 \citep{Buder2021GALAHSurveyThird} for this comparison. The 1D-NLTE $A(\rm{Li})$ values are determined using the Spectroscopy Made Easy \citep[SME,][]{Valenti1996SpectroscopyMadeEasy, Piskunov2017SpectroscopyMadeEasy} code with 1D \textsc{MARCS} \citep{Gustafsson2008GridMARCSModel} model atmospheres, offering a consistent baseline for comparison. We perform a direct comparison with these data to validate our own LTE results. To ensure the reliability of the data, we select only stars that meet the following commonly used quality criteria:
\begin{enumerate}
    \item $\texttt{flag\_a\_li} = 0$ and exclude \ALi\ of upper limits;
    \item $\texttt{snr\_px\_ccd3} > 30$ and $\texttt{flag\_sp} = 0$ for overall spectrum quality;
    \item $\texttt{flag\_red} = 0$ and $\texttt{nn\_flag\_li\_fe} = 0$ for reduction reliability and abundance.
\end{enumerate}

After applying these criteria, we obtain 216 common stars and recalculate their $A\rm(Li)_{LAMOST}$ using the stellar parameters provided by GALAH and visually inspected each spectrum. After discarding poor-quality fits and excluding sources with $A\rm(Li)_{LAMOST}<1.5\,dex$ to avoid unreliable detections, 62 stars are retained for comparison.

Figure~\ref{fig:vsGALAH} illustrates the comparison between our derived $A\rm(Li)_{LAMOST}$ and the GALAH DR3 values. The results exhibit good consistency across the abundance range of $1.5<A\rm(Li)<3.5\,dex$. For the 62 common stars, we find a systematic offset of approximately 0.20\,dex, which tends to increase at higher $A\rm(Li)$ levels. This discrepancy can be attributed primarily to the departures from LTE reported by \citet{Lind2009DeparturesLTENeutrala}, where the NLTE corrections become more negative with increasing abundance in Li-rich giants. Nevertheless, this discrepancy has little effect on our identification of Li-rich giants, as these objects are defined by abundances that exceed roughly $A\rm(Li) \sim 1.5$\,dex, where the difference in Li abundances is small.

\subsection{Selection Refinement}
Following the validation described in Section~\ref{subsec:compare}, it becomes necessary to adopt appropriate criteria to identify Li-rich giants from our sample set.

The lithium abundance threshold commonly used to distinguish Li-rich giants from Li-normal giants is \ALi\ $> 1.5$\,dex \citep[e.g.,][]{Brown1989SearchLithiumrichGiant, Kumar2011OriginLithiumEnrichment}. Among the 615,426 LRS giants for which \ALi\ has been derived, applying this criterion results in 66,857 stars with \ALi\ $> 1.5$\,dex, even after excluding spectra with \snrr\ $< 20$ and upper limits. As discussed in Section~\ref{subsec:limit}, low spectral quality and line-blending can substantially affect the derived abundances, which requires further refinement of our selection. Compared with the results based on high-resolution spectra (Figure~\ref{fig:vsHRSlte}), it can be seen that our results tend to overestimate \ALi\ by approximately 0.15\,dex. Taking this offset into account, we adopt a revised threshold of \ALi\ $> 1.65$\,dex to identify Li-rich giants. It is necessary to exclude potential binaries from our catalog. We therefore cross-match our samples with spectroscopic binary catalogs \citep{Zheng2023SearchingDoublelineSpectroscopic, Kovalev2024Detection12426, Li2025MiningDoublelineSpectroscopic} and remove spectroscopic binary candidates. 

Consequently, after these refinements, 38,333 stars remain with \ALi\ $> 1.65$\,dex. These candidates are then visually inspected to remove contamination and retain only the most reliable Li-rich giants for subsequent analysis.

\subsection{Manual Inspection} \label{subsec:manual}

To improve the reliability of the selected Li-rich candidates, we carefully re-examine all the spectra through manual inspection. For this purpose, we develop a web-based spectral viewer that allows interactive visualization of the observed spectra and their corresponding synthetic templates. Each spectrum is displayed in the wavelength range of 6670––6740\,\AA, with the $6708\pm6$\,\AA\ \ion{Li}{1} feature highlighted. Synthetic spectra with \ALi=1.0––4.0\,dex in 0.5\,dex steps are plotted together with the observed spectrum to indicate the strength of the \ion{Li}{1} line.

\begin{figure}[!htbp]
	\centering
	\fig{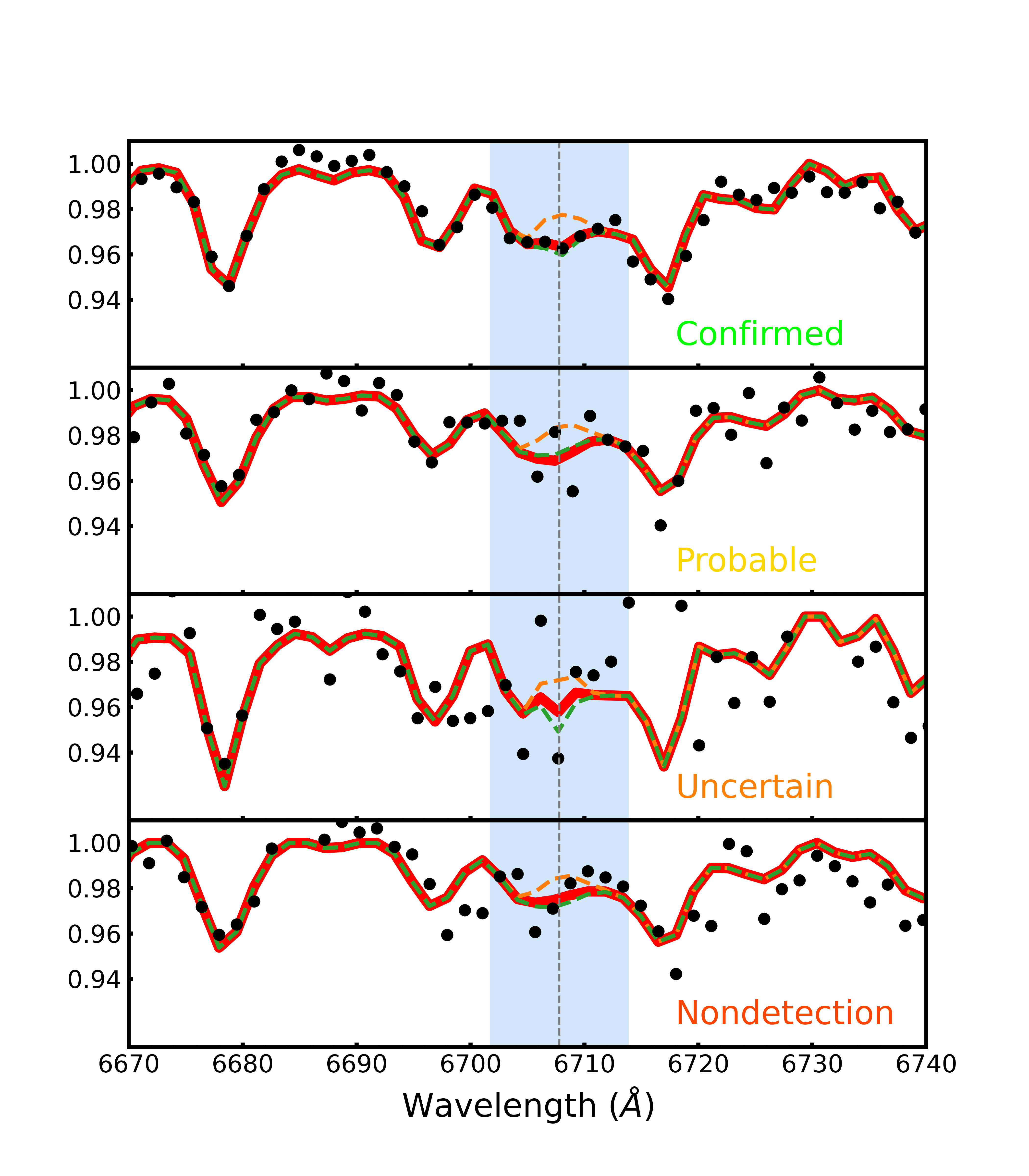}{1\columnwidth}{}
	\caption{Representative examples of manually inspected spectra with typical stellar parameters (\teff\ $\sim 5000$\,K, \logg\ $\sim 2.5$\,dex, \feh\ $\sim -0.1$\,dex) and high spectral quality (\snrr\ $> 50$) for the four classification categories. The orange and green dashed curves correspond to comparison templates with \ALi\ $=$ 1.5\,dex and 2.0\,dex, respectively}
	\label{fig:eyecheck_samples}
\end{figure}

The inspection aims to verify whether the \ion{Li}{1} absorption is intrinsically or artificially enhanced by noise. In some cases, a spuriously high \ALi\ value could result from line blending, strong noise, or an inaccurate local continuum, which may distort the fitting. Additional effects such as cosmic rays, bad pixels, or other artifacts are also recognized, as they can mimic the appearance of a \ion{Li}{1} feature. Manual inspection is therefore essential to identify these cases and retain only spectra with reliable Li detections.

\begin{figure}[!htbp]
	\centering
	\fig{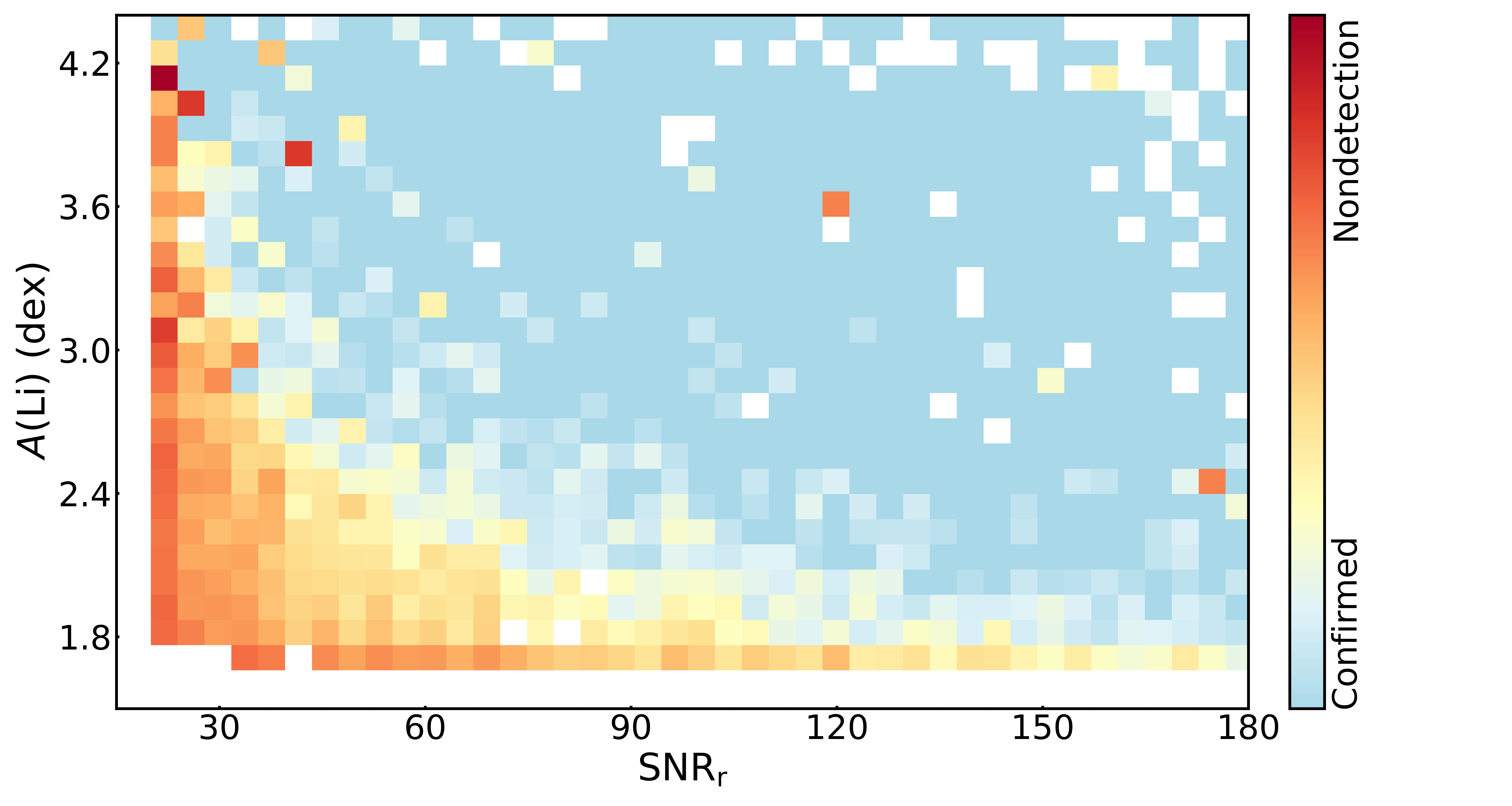}{1\columnwidth}{}
	\caption{The distribution of the manually inspected candidates, color-coded by the mean classification score.}
	\label{fig:snr_vs_ali}
\end{figure}

Figure~\ref{fig:eyecheck_samples} presents representative examples of the four classification categories used in visual inspection: \textit{Confirmed}, \textit{Probable}, \textit{Uncertain}, and \textit{Nondetection}, Spectra marked as \textit{Confirmed} show a clear \ion{Li}{1} characteristic or strong absorption that exceeds the local noise level. \textit{Probable} cases are somewhat affected by noise or blending, but still display a plausible \ion{Li}{1} signal. \textit{Uncertain} spectra have \ion{Li}{1} lines so weak that they are difficult to distinguish from noise, whereas \textit{nondetection} cases are dominated by noise, blending or no recognizable \ion{Li}{1} absorption. During classification, auxiliary flags, such as \texttt{flag\_Halpha}, \texttt{flag\_feature}, and \texttt{flag\_SNR}, are also recorded. To ensure classification consistency, the entire candidate set is reviewed multiple times.

Figure~\ref{fig:snr_vs_ali} shows the distribution of candidates manually inspected across the space of \ALi---$\rm S/N_r$. Manual inspection reveals a clear separation between reliable and unreliable detections: \textit{confirmed} and \textit{probable} Li-rich giants distribute toward higher \ALi\ and \snrr\ regions, whereas \textit{uncertain} and \textit{nondetection} samples are concentrated in the lowest \snrr\ and \ALi\ value region.

\begin{deluxetable}{lrr}
\tabletypesize{\small}
\tablewidth{0pt}
\tablecaption{Classification of Li-rich Giant Candidates \label{tab:classification}}
\tablehead{
\colhead{Category} & \colhead{Sample Size} & \colhead{Fraction (\%)} 
}
\startdata
Total         & 38,333 & 4.80 \\
\hline
Confirmed     & 13,785 & 1.73 \\
Probable      & 6633  & 0.83 \\
Uncertain     & 7404  & 0.93 \\
Nondetection & 10,482 & 1.31 \\
\enddata
\end{deluxetable}

Table~\ref{tab:classification} summarizes the results of the manual inspection. Following the refined criteria of 1.65\,dex, 38,333 candidates are selected for visual inspection, among which we identified 13,785 \textit{confirmed} (1.73\%) and 6633 \textit{probable} (0.83\%) Li-rich giants. This results in a total of 20,418 positive detections, about 50\% of the initial candidates and 2.5\% of all giants of LAMOST DR9. This fraction is consistent with previous studies, which typically report a Li-rich giant occurrence rate of 0.5––2.5\% \citep[e.g.,][]{Gonzalez2009LirichGiantsGalactic, Monaco2011LithiumrichGiantsGalactic, Kumar2011OriginLithiumEnrichment, Lebzelter2012LithiumAbundancesRed, Liu2014LithiumAbundancesLargea, Deepak2019StudyLithiumRich, Charbonnel2020LithiumRedGiant} when adopting a threshold of \ALi\ $> 1.5$\,dex. The remaining spectra are classified as \textit{uncertain} (0.93\%) or \textit{nondetection} (1.31\%), mainly due to low spectral quality or ambiguous line features. Our manual inspection substantially reduces false positives and produces a reliable sample of Li-rich giants for subsequent analysis.

\section{Discussion} \label{sec:dis}

\subsection{The Final Li-rich Catalog}

Following the manual inspection described above, a final catalog of Li-rich giants is compiled from LAMOST DR9. To assess the coverage of the detections, a comparison is made with the catalog of \citet{Gao2019LithiumrichGiantsLAMOST}, which identified 10,500 Li-rich giants from the LAMOST LRS DR7. Among the DR7 candidates, 6735 objects are cross-matched, of which 94\% are confirmed to be Li rich in this work, while the remaining 3766 objects generally exhibit \ALi\ $< 1.65$\,dex. The final catalog contains 20,400 Li-rich giants, nearly twice the number reported by \citet{Gao2019LithiumrichGiantsLAMOST}. After removing the overlapping sources, 67\% of these objects are newly identified, reflecting improved coverage achieved with the expanded DR9 dataset.

\begin{deluxetable}{lcc}
\tablecaption{Description of the Li-rich Giant Catalog}\label{tab:catalog}
\tablehead{
\colhead{Column} & \colhead{Unit} & \colhead{Description}
}
\startdata
DESIG                      & --- & LAMOST designation \\
Source\_id                 & --- & Unique \textit{Gaia} Identifier \\
LMJD                       & --- & Local Modified Julian Day \\
plainID                    & --- & LAMOST Plan ID in use \\
spID                       & --- & LAMOST Spectrograph ID \\
fiberID                    & --- & LAMOST Fiber ID of Object \\
RA                         & deg & Right ascension (J2000) \\
Decl.                      & deg & Decl. (J2000) \\
SNR\_r                     & --- & Signal-to-noise of r band \\
\teff                      & K   & Effective temperature \\
$\sigma_{T_{\rm eff}}$     & K   & Uncertainty of effective temperature \\
\logg                      & dex & Surface gravity \\
$\sigma_{\log g}$          & dex & Uncertainty of surface gravity \\
\feh                       & dex & Metallicity \\
$\sigma_{\rm [Fe/H]}$      & dex & Uncertainty of metallicity \\
RV                         & $\mathrm{km \cdot s^{-1}}$ & Radial velocity \\
$\sigma_{\rm RV}$          & $\mathrm{km \cdot s^{-1}}$ & Uncertainty of radial velocity \\
\ALi                       & dex & Li abundance \\
\ALi\_lower                & dex & Lower limit of Li abundance \\
\ALi\_upper                & dex & Upper limit of Li abundance \\
$\sigma_{A(\mathrm{Li})\_T_{\mathrm{eff}}}$& dex & Uncertainty of \ALi\ due to $\sigma_{T_{\rm eff}}$ \\
Flag\_Li-rich              & --- & 1---Confirmed, 2---Probable\\
Flag\_H$\alpha$            & --- & H$\alpha$ emission flag \\
Flag\_Feature              & --- & Local spectral feature flag \\
Flag\_SNR                  & --- & Signal-to-noise quality flag \\
\enddata 
\tablecomments{The full machine-readable table is available online.}
\end{deluxetable}

Table~\ref{tab:catalog} summarizes the structure of the final catalog of Li-rich giants. It includes basic source information such as the LAMOST identities, \textit{Gaia} source identifiers, equatorial coordinates, the stellar parameters (\teff, \logg, \feh), RVs with their corresponding uncertainties. For each object, we provide the measured lithium abundance, the associated lower and upper limits, and the uncertainty in \ALi\ induced by \teff. Additionally, several auxiliary flags record the results of manual inspection, including the classification categories and the auxiliary quality flags. These auxiliary flags are manually recorded in binary format (1/0). Specifically, \texttt{flag\_Halpha} is assigned to stars exhibiting H$\alpha$ emission or significant infilling to distinguish them from the majority of Li-rich giants, as such features might indicate distinct enrichment channels or contamination. We performed a preliminary comparison of the \ALi\ distributions for Li-rich giants with \texttt{flag\_Halpha=1} and \texttt{flag\_Halpha=0}, and no significant difference is found.

\begin{figure}[!htbp]
	\centering
	\fig{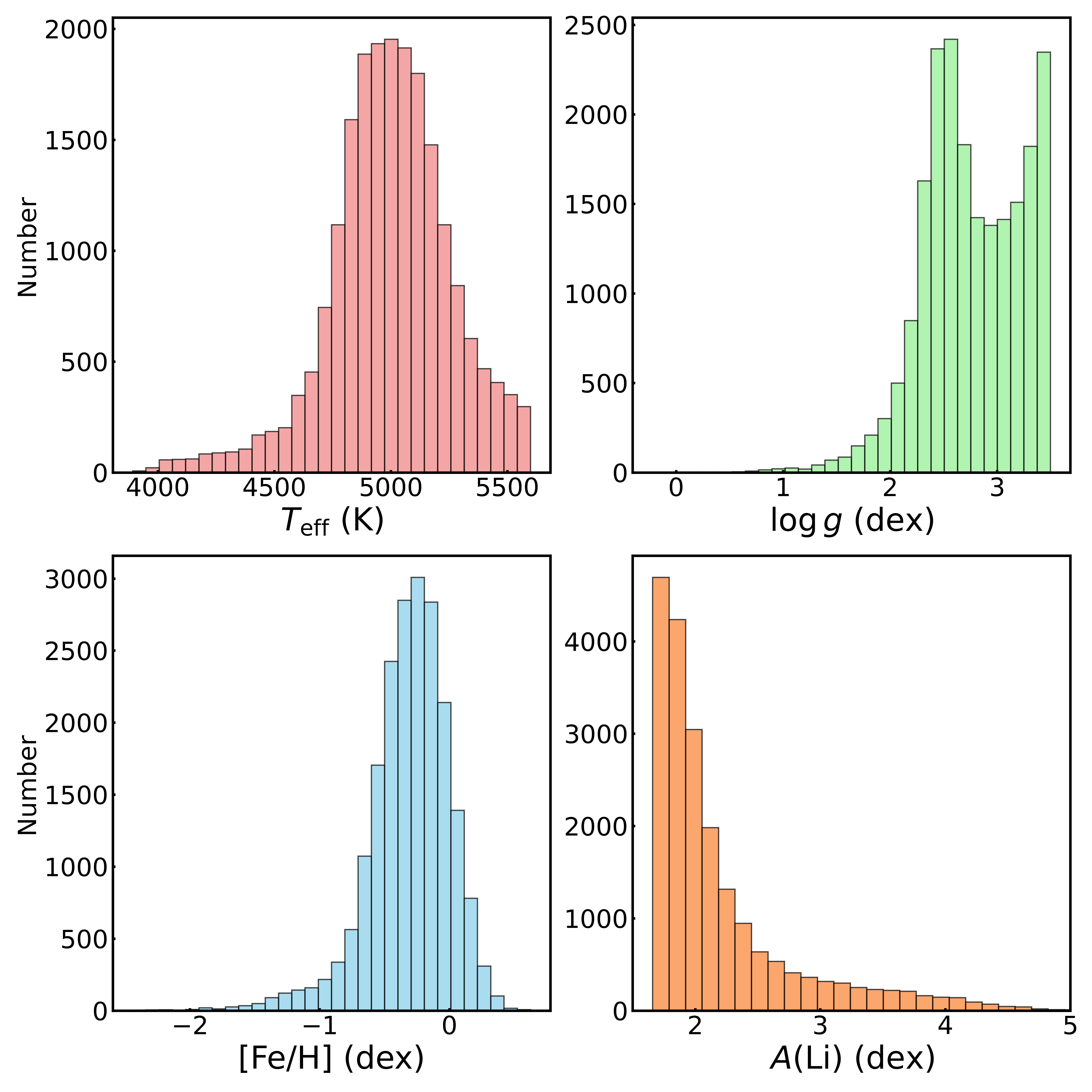}{1\columnwidth}{}
	\caption{Histograms of the stellar parameters (\teff, \logg, \feh), and \ALi\ for the Li-rich giant sample.}
	\label{fig:Li_rich_Hist}
\end{figure}

Figure~\ref{fig:Li_rich_Hist} presents the distributions of the stellar parameters for our Li-rich giants. The \teff\ peaks around 5000~K, and the \logg\ have two peaks near 3.5\,dex and 2.5\,dex. The \feh\ are concentrated around $-0.25$\,dex. The lithium abundances show a rapid decay from the selection threshold, which is consistent with previous studies.

\subsection{Error Estimation}

As discussed in previous studies, the intrinsic uncertainty of \ALi\ is mainly influenced by the spectral quality and tends to increase with increasing \teff\ and \logg. We estimate the upper and lower limits for the Li-rich giants based on our $\chi^2$ minimization method. The $1\sigma$ confidence interval of the $\chi^2$ curve is adopted to quantify the corresponding uncertainty.

\begin{figure}[!htbp]
	\centering
	\fig{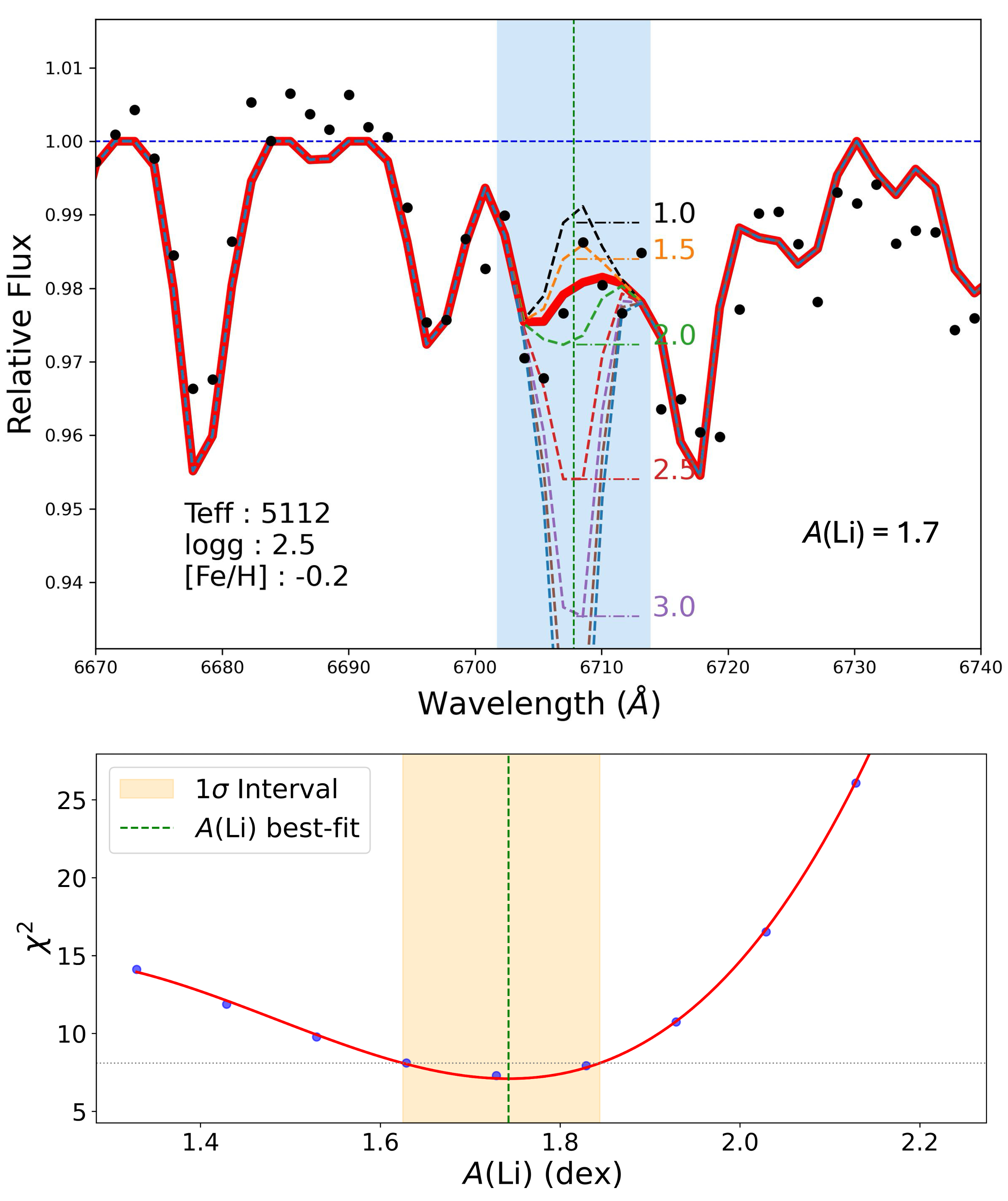}{0.9\columnwidth}{}
	\caption{Top panel: a representative observed spectrum (black) with best-fit template (red line) and comparison templates with different \ALi\ values (dashed lines). Bottom panel: the corresponding $\chi^2$ variation with a third-order polynomial fit (red curve) and the $1\sigma$ confidence region indicated.}
	\label{fig:Chi2}
\end{figure}

As illustrated in Figure~\ref{fig:Chi2}, the upper panel presents a representative spectrum with \ALi\ near the adopted threshold, along with template spectra at different \ALi\ values. The lower panel displays the variation of $\chi^2$ used to determine the confidence region of $1\sigma$. This once again demonstrates that when noise, line blending, or continuum errors are present, the sensitivity of minimization $\chi^2$ decreases, leading to larger uncertainties in the derived \ALi.

Additionally, the uncertainty in \teff\ is a dominant source of error in \ALi\ measurement \citep[e.g.,][]{Kirby2012DiscoverySuperLiRich, Wang20243DNLTELithiuma, Zhou2025LirichUnevolvedStars}. To quantify this impact, we re-calculated the lithium abundances by perturbing the \teff\ with the individual errors $\pm \sigma_{T_{\text{eff}}}$ provided by LASP. In cases where a perturbed parameter might fall outside our pre-computed model grid, the uncertainty is estimated using the deviation from the single valid side. Seven stars in our sample fall outside the valid range for both perturbations and are assigned as \texttt{NaN}. For the majority of our Li-rich giants, the \teff-induced uncertainty $\sigma_{A(\mathrm{Li})\_T_{\mathrm{eff}}}$ is less than 0.1\,dex, only a few targets with large \teff\ uncertainty ($> 400$\,K) exhibit errors of $\sim 0.5$\,dex.

\subsection{Proportion of Li-rich Giants}
The occurrence rate of Li-rich giants provides insight into the conditions under which lithium enrichment can be observed. We evaluate the fraction of Li-rich stars in bins of effective temperature, surface gravity, and metallicity relative to all giants in the same parameter range.

\begin{figure}[!htbp]
	\centering
	\fig{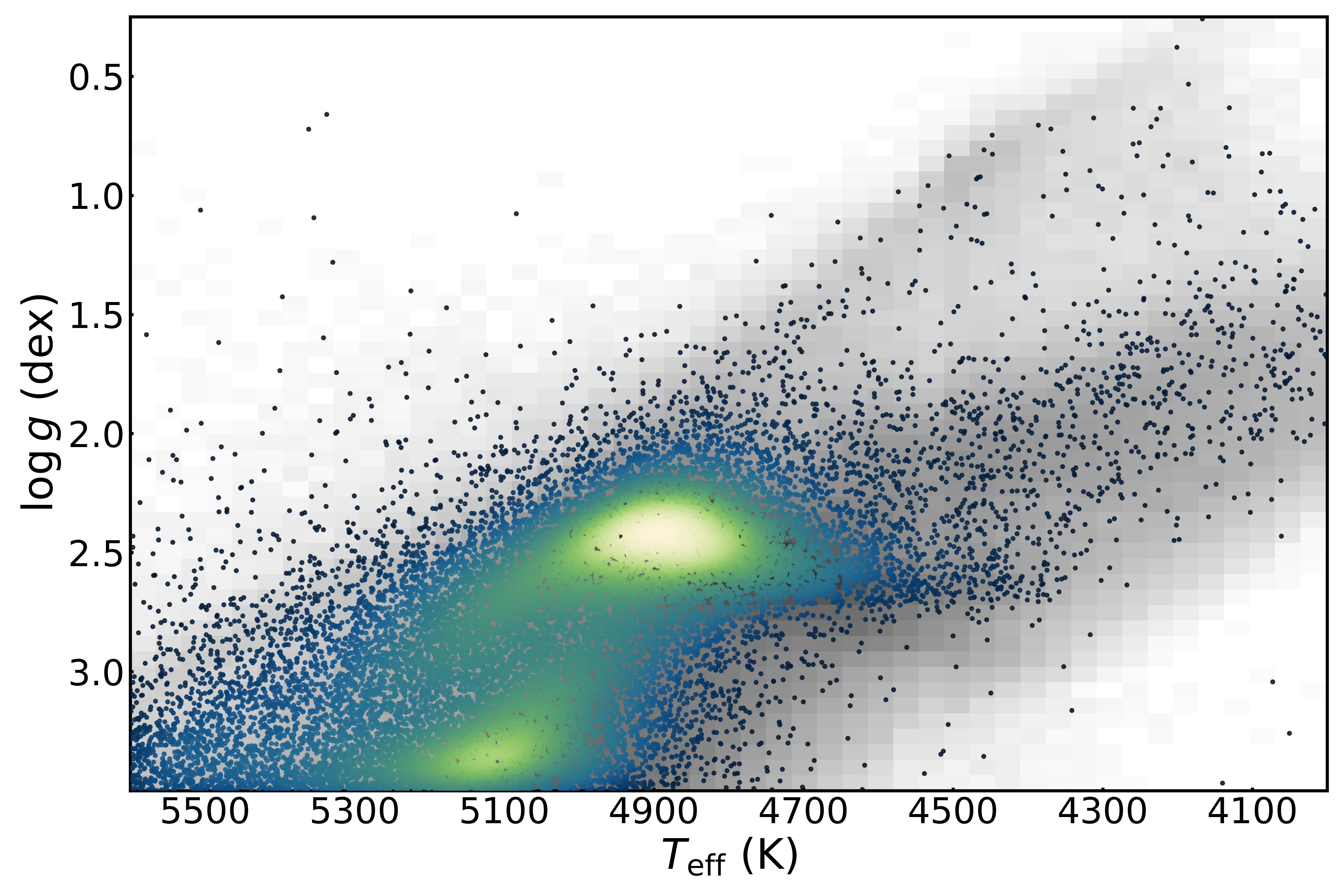}{1\columnwidth}{}
	\caption{HR diagram of LAMOST DR9 giants (gray) and identified Li-rich giants (color-coded by density).}
	\label{fig:Teff_logg_proportion1}
\end{figure}

\begin{figure}[!htbp]
	\centering
    \fig{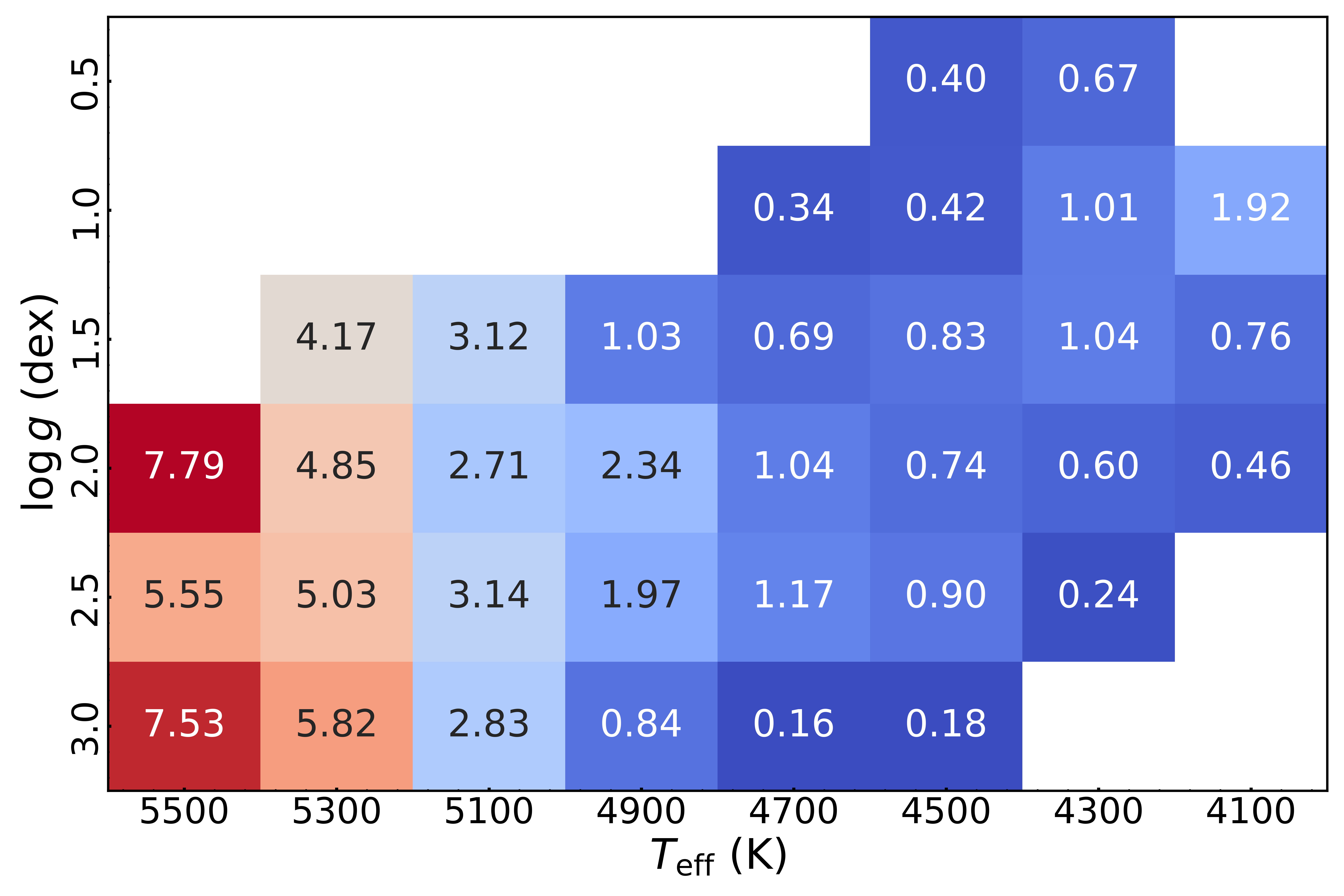}{1\columnwidth}{}
	\caption{Proportion of Li-rich giants across the \teff––\logg\ plane. The percentages are noted in each grid.}
	\label{fig:Teff_logg_proportion2}
\end{figure}

As shown in Figure~\ref{fig:Teff_logg_proportion1}, Li-rich giants form a prominent peak in the RC region, suggesting a noticeable enhancement of $\text{Li}$ during the HeB phase. In addition, a secondary enhancement at higher $\log g \rm > 3.0\,dex$ likely corresponds to the RGB bump. This feature arises from a temporary evolutionary stall due to a discontinuity in hydrogen abundance left by deep convection, after this stage surface lithium could be rapidly depleted by mixing processes. In Figure~\ref{fig:Teff_logg_proportion2}, a noticeable enhancement appears towards the higher \teff\ and \logg. Within the temperature range of 5000––5600\,K, the hotter stars exhibit a higher Li-rich proportion. This feature might be associated with the relatively larger internal uncertainties in this parameter region where the \ion{Li}{1} line becomes weaker and more sensitive to spectral noise, as discussed in Section~\ref{subsec:manual}. For the rest of the giants, no clear trend can be discerned from the heatmap. A finer analysis will require more detailed information on their evolutionary stages.

\begin{figure}[!htbp]
	\centering
	\fig{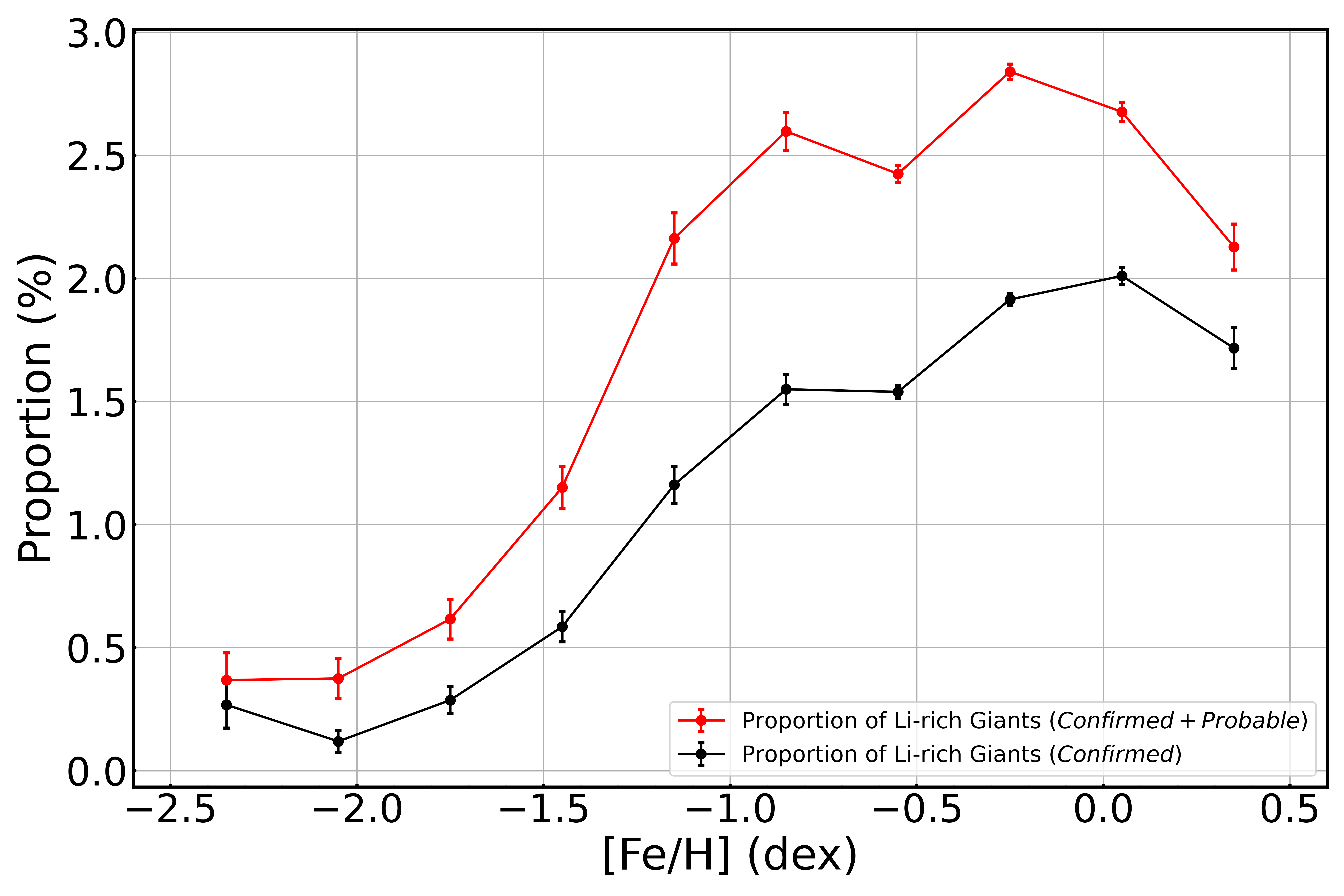}{1\columnwidth}{}
	\caption{Proportion of Li-rich giants as a function of \feh, binned in 0.3\,dex intervals. The red curve denotes full catalog of our Li-rich giants (\textit{Confirmed $+$ Probable}), while the black curve represent only the samples which marked as \textit{Confirmed}.}
	\label{fig:Fe_H_proportion}
\end{figure}

The dependence of the Li-rich fraction on metallicity is illustrated in Figure~\ref{fig:Fe_H_proportion}, where the proportion generally increases with \feh\ from $-$2.5 to 0.0\,dex and shows a slight decrease beyond solar metallicity. This observed pattern may reflect the evolutionary imprints of Li enrichment, while, at higher metallicities, some mechanisms (such as stronger dilution or envelope mixing) may suppress the observable abundance.

\begin{figure}[!htbp]
	\centering
	\fig{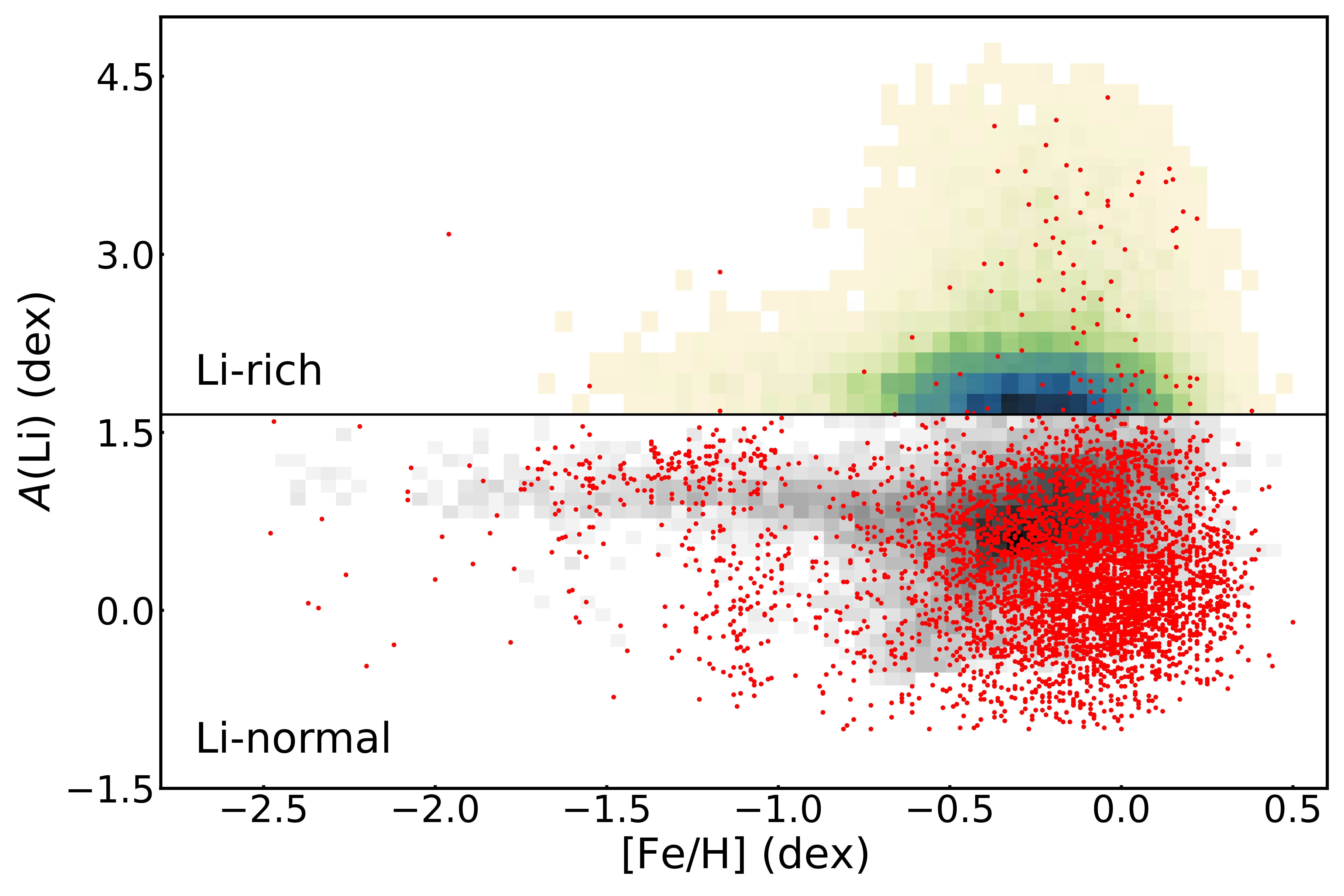}{1\columnwidth}{}
	\caption{The distribution of A(Li) as a function of [Fe/H]. The color-coded density represents the Li-rich giants identified in this work. For comparison, the underlying gray-scale density and red dots are collected from the GALAH and \textit{Gaia}-ESO surveys, respectively.}
	\label{fig:ALi_Fe_H}
\end{figure}

Figure~\ref{fig:ALi_Fe_H} presents the distribution of \ALi\ versus \feh\ for our identified Li-rich giants. To provide a comparative baseline, we utilize the high-resolution GALAH \citep{Buder2021GALAHSurveyThird} and \textit{Gaia}-ESO \citep{Franciosini2022GaiaESOSurveyLithiuma} surveys, selected after applying quality flags, upper limit exclusions, and SNR controls. We note that the apparent discontinuity between the Li-rich and Li-normal populations arises from our resolution constraints ($R \sim 1800$), as well as differences in data sources and sample sizes across the surveys. As illustrated in Figure~\ref{fig:ALi_Fe_H}, in the metal-poor regime, numerous Li-normal giants form a plateau at \ALi\ $\sim$ 1.0\,dex, where Li-rich detections are relatively rare. Additionally, a few stars in our sample exhibit notably high lithium abundances, appearing primarily clustered within the metallicity range of $\rm -0.5< \feh <0$\,dex.

Overall, the preliminary analysis presented here provides a statistical view of the Li-rich fraction across fundamental stellar parameters. Future work will conduct more detailed investigations of the enrichment mechanisms and evolutionary features of the Li-rich giants identified in this study, particularly with the ongoing and expanding LAMOST survey.

\section{Summary} \label{sec:con}
In this work, we conduct a systematic search for Li-rich giants using the extensive dataset from the LAMOST low-resolution survey. A template matching method is developed to measure the \ALi\ from the \ion{Li}{1} 6708\,\AA\ absorption line with templates generated with the ATLAS9 model. The derived lithium abundances are validated through comparison with previous high-resolution and independent measurements, showing good agreement with a deviation of $\sim$0.15\,dex. Adopting a selection threshold of $A(\mathrm{Li}) > 1.65$\,dex, we identify a total of 20,418 Li-rich giants from the LAMOST DR9 dataset after a thorough visual inspection. It is important to stress that visual inspection significantly reduces false positives in our original samples, and the final catalog corresponds to approximately 2.5\% of all giants, representing one of the largest and most homogeneous samples of Li-rich giants available.

We further investigate the occurrence rate of Li-rich giants across various stellar parameters, which can provide important observational constraints for understanding the evolutionary stages and mechanisms for lithium enrichment in evolved stars.

The catalog and derived lithium abundances presented in this work provide a robust foundation for future investigations into the origin and evolution of lithium in the Galaxy. In particular, high-resolution spectroscopic follow-up of the most extreme sources—such as super Li-rich giants and metal-poor candidates—will be crucial for constraining their detailed chemical compositions and evolutionary histories, thereby offering new insights into the mechanisms driving lithium enrichment in evolved stars.

\begin{acknowledgments}
This work is supported by the Strategic Priority Research Program of Chinese Academy of Sciences, grant No. XDB1160101. This research is also supported by the National Natural Science Foundation of China under Grant Nos. 12090040/4, 12373036, 12022304, 12222305, 12427804, 12573089, the National Key R\&D Program of China Nos. 2019YFA0405502, 2024YFA1611601, the Scientific Instrument Developing Project of the Chinese Academy of Sciences, Grant No. ZDKYYQ20220009, and the International Partnership Program of the Chinese Academy of Sciences, Grant No. 178GJHZ2022047GC. H.-L.Y. acknowledges the support from the Youth Innovation Promotion Association, Chinese Academy of Sciences. We acknowledge the science research grants from the China Manned Space Project with NO.CMS-CSST-2021-B05. This work is supported by Chinese Academy of Sciences President's International Fellowship Initiative. Grant No. 2020VMA0033. (id. 2019060). J. C. is supported by the Postdoctoral Fellowship Program of CPSF under grant No. GZC20232780. Guoshoujing Telescope (the Large Sky Area Multi-Object Fiber Spectroscopic Telescope LAMOST) is a National Major Scientific Project built by the Chinese Academy of Sciences. Funding for the project has been provided by the National Development and Reform Commission. LAMOST is operated and managed by the National Astronomical Observatories, Chinese Academy of Sciences.
\end{acknowledgments}

\bibliography{sample631}{}
\bibliographystyle{aasjournal}

\end{document}